\documentclass[a4paper,fleqn]{cas-sc}

\usepackage[numbers,sort&compress]{natbib}
\usepackage{mathtools}

\usepackage{tikz}
\usetikzlibrary{positioning,fit,arrows.meta,backgrounds,calc}
\usepackage{subcaption}
\usepackage{placeins}
\usepackage{pifont}
\usepackage{enumitem}
\IfFileExists{twemojis.sty}{\usepackage{twemojis}}{\newcommand{\twemoji}[1]{\rule{0.35cm}{0.35cm}}}

\begin{document}
\let\WriteBookmarks\relax
\def\floatpagepagefraction{1}
\def\textpagefraction{.001}

\shorttitle{Can the Cloud Drive?}
\shortauthors{Parsa et~al.}

\title[mode=title]{Can the Cloud Drive? Infrastructure Feasibility of Offloading Autonomous Driving Across 5G and 6G}

\author[1]{Pouya Parsa}
\ead{parsa025@umn.edu}
\credit{Conceptualization, Investigation, Methodology, Software, Writing}

\author[2]{Kawon Han}
\ead{kawon.han@unist.ac.kr}
\credit{Conceptualization, Investigation, Methodology, Writing}

\author[1]{Seongjin Choi}[orcid=0000-0002-6862-3091]
\cormark[1]
\ead{chois@umn.edu}
\credit{Conceptualization, Investigation, Methodology, Project administration, Supervision, Validation, Visualization, Writing}

\affiliation[1]{organization={Department of Civil, Environmental, and Geo-Engineering, University of Minnesota, Twin Cities},
            city={Minneapolis},
            postcode={55455},
            state={Minnesota},
            country={United States}}

\affiliation[2]{organization={Department of Electrical Engineering, Ulsan National Institute of Science and Technology (UNIST)},
            city={Ulsan},
            postcode={44919},
            country={South Korea}}

\cortext[1]{Corresponding author}

\begin{abstract}
Frontier autonomous-driving models---especially vision-language-action (VLA) models, whose forward pass approaches $\sim$60~TFLOPs---are outgrowing economical onboard deployment, since peak hardware sits idle most of the day.
Cloud inference can instead share GPUs across active vehicles, but the vehicle must upload through a capacity-limited uplink, reach a GPU without queueing, and return a decision within the closed-loop budget.
This paper asks: \textit{can the cloud drive?}
We answer with an analytical framework coupling communication limits, a roofline GPU service model, stochastic latency, and utilization-aware cost across three model classes, three offloading strategies, and three communication generations, applied to New York City.
Separating a reactive 100~ms budget from a 300~ms deliberative tier (presuming an onboard reactive fallback), we find three \emph{nested} binding regimes.
\emph{Communication} binds first in dense cells: 5G fails early, 5G-Advanced is the practical threshold for feature-level offloading, and 6G adds headroom.
\emph{Compute} binds next under the reactive budget: near-term VLA is latency-infeasible regardless of bandwidth, because autoregressive FP16 decode is memory-bandwidth-bound ($\sim$114~ms on 2025 hardware). Its floor clears 100~ms around 2027; 6G then admits feature-level VLA by $\sim$2028, 5G-Advanced only at light loading and not the dense corridor, and the deliberative tier from 2026.
\emph{Cost} binds last: once admissible, utilization-pooled cloud GPUs undercut onboard hardware for VLA, whose baseline (up to \$8{,}500 per vehicle-year) is expensive and idle; feature-level offloading (S2) is where the VLA cost crossover concentrates.
Latency decides which model is admissible in which year; cost decides whether it is economical.
\end{abstract}

\begin{highlights}
\item Framework links AV model complexity, 5G/6G limits, and cloud GPU economics
\item Three nested NYC regimes: communication binds, then VLA compute, then cost
\item Feature-level offloading (S2) is where the VLA cloud cost crossover concentrates
\item Latency sets which model is admissible each year; cost sets if it is economical
\end{highlights}


\maketitle

\section{Introduction}
\label{sec:introduction}

Autonomous driving (AD) is moving from modular perception-prediction-planning pipelines toward end-to-end models that map sensor inputs directly to driving actions. This transition has widened the computational range of AD models substantially.
Compact end-to-end (E2E) architectures such as UniAD~\citep{hu2023uniad} and VAD~\citep{jiang2023vad} operate around a few TFLOPs (tera floating-point operations) per forward pass; Vision Language Model (VLM)-based models such as DriveLM~\citep{sima2024drivelm} add language-conditioned reasoning; and recent Vision-Language-Action (VLA) models such as EMMA~\citep{hwang2025emma} and Alpamayo~\citep{nvidia2026alpamayo} push perception, reasoning, and action into a single large model.
While current automotive processors can marginally accommodate E2E and VLM workloads, running larger models such as VLAs remains a practical deployment challenge.

The straightforward response is to put more hardware in every vehicle.
For personal vehicles, however, this approach is economically inefficient since each vehicle has to be equipped with hardware sized for its peak inference load.
%
The peak arises only during rare driving moments when the full autonomous-driving pipeline must run end-to-end, yet the same hardware sits idle for the roughly 95\% of the day the vehicle is parked~\citep{doe2024fotw1356}.
%
%
Cloud or edge inference changes this cost structure by sharing expensive hardware---especially graphics processing units (GPUs)---across the vehicles active at any moment.
It also simplifies fleet-wide model updates and allows computational capacity to scale with actual demand rather than with the total number of vehicles sold.
These advantages motivate the central question of this paper: \textit{can the cloud drive?}

The difficulty is that cloud driving is not constrained by cloud compute alone. It couples three systems usually analyzed separately: \textbf{(1) the driving-model split}, which determines which stages of the AD pipeline---sensing, perception, planning, and decoding---remain onboard and which are offloaded to the cloud; \textbf{(2) vehicle-to-network communication}, which determines how many active vehicles a single cell can admit while each continuously uploads sensor data or features to the cloud; and \textbf{(3) the cloud GPU service system}, which determines whether shared GPUs can serve many vehicles' bursty requests within the closed-loop deadline that all three systems must jointly satisfy---a reactive 100~ms budget for immediate control, or a relaxed 300~ms deliberative budget that is admissible only behind an onboard reactive fallback.
Each of these three systems has been studied by its own literature, but rarely as parts of one deployment problem. 
Driving-model literature characterizes perception, planning, and reasoning quality, but generally treats inference as a vehicle-side workload~\citep{chen2024e2eadsurvey,hu2023uniad,jiang2023vad,sima2024drivelm,hwang2025emma,liu2019edgedriving}.
Vehicular communication literature characterizes DSRC, C-V2X, 5G, 5G-Advanced, and 6G links, but rarely translates those link capabilities into sustained Autonomous Vehicle (AV) uplink loads for cloud inference~\citep{naik2019ieee80211bd,garcia2021tutorial,lin2022overview5gadv,chen20235gadvanced,noor2022sixgv2x}.
Vehicular edge computing studies optimize task offloading, latency, energy, or resource allocation, but often treat the workload as a generic application task rather than as an E2E, VLM, or VLA driving pipeline with different split points and residual onboard hardware~\citep{desouza2020computationoffloading,ahmed2022vehiculartaskoffloading,karimi2022taskoffloading}.
What is missing is a single framework that asks the deployment question in the order an operator would face it: first, whether cloud offloading can meet communication and latency requirements; second, whether feasible cloud offloading is more economical than equipping every vehicle with the full onboard hardware.

To address the gap, we develop a unified framework and apply it to a single New York City case study, read through three \emph{nested} binding regimes.
Dense urban cells first make uplink capacity the binding constraint (the \emph{communication-bound} regime).
For branches that clear the uplink, the reactive control budget then makes near-term VLA inference latency-infeasible regardless of bandwidth, because autoregressive decoding is memory-bandwidth-bound (the \emph{compute-bound} regime).
Only once a branch is both communication- and latency-admissible does the comparison become economic, governed by how well utilization lets shared cloud GPUs undercut per-vehicle onboard hardware (the \emph{cost-bound} regime).
Using one dense metropolitan setting for all three regimes holds fleet size, cell density, and cost aggregation fixed, so the regimes are separated by which constraint binds rather than by changing the scenario.
Our contributions are:
%
\begin{enumerate}[leftmargin=*]
  \item \textbf{Three-system framework.} A unified analytical framework that links the driving-model split (system 1) to communication-generation limits (system 2) and stochastic cloud GPU service (system 3), integrating a roofline inference model for autoregressive VLM/VLA decoding, a GPU evolution model calibrated to A100$\to$B300 trends, and edge infrastructure optimization formulated as capacitated facility location under tail-latency constraints.
  \item \textbf{Three-regime case study.} A single New York City case study evaluated under a reactive 100~ms and a deliberative 300~ms budget, separating three nested binding regimes --- communication, compute (memory-bandwidth-bound VLA decode), and cost --- so that latency determines which model class is admissible in which year and cost determines whether it is economical.
  \item \textbf{Policy implications.} Recommendations for uplink-heavy spectrum allocation, shared edge infrastructure co-investment, cloud-dependent safety certification with local fallback, and utilization-aware infrastructure sizing.
\end{enumerate}

\section{Related Work}
\label{sec:related_work}

We organize related work into three clusters that map directly to the three systems framed in Section~\ref{sec:introduction}: autonomous-driving models (system~1), vehicular communication (system~2), and vehicular edge computing (system~3). Each cluster covers a piece of the cloud-driving deployment question, but none integrates the three.

\subsection{Driving models (E2E, VLM, VLA).} This literature explains why cloud deployment is now worth asking about: model classes that were once comfortably onboard are migrating toward inference budgets that current automotive processors cannot economically support. The first wave is E2E models such as UniAD~\citep{hu2023uniad}, VAD~\citep{jiang2023vad}, ParaDrive~\citep{weng2024paradrive}, and TransFuser~\citep{chitta2023transfuser}, which fuse perception, prediction, and planning into a single forward pass at a few TFLOPs. This is a workload that current automotive system-on-chip (SoC) hardware can support. A second wave adds language-conditioned reasoning: DriveLM~\citep{sima2024drivelm}, DriveVLM~\citep{tian2024drivevlm}, LMDrive~\citep{shao2024lmdrive}, and DriveGPT4~\citep{xu2024drivegpt4} use vision-language backbones to connect driving scenes with textual explanations or instructions, which adds substantial compute and memory pressure beyond the E2E baseline. The most recent wave is vision-language-action (VLA) models~\citep{jiang2025surveyvla} such as EMMA~\citep{hwang2025emma}, OpenVLA~\citep{kim2024openvla}, and Alpamayo~\citep{nvidia2026alpamayo}, which merge perception, prediction, and planning into a single large multimodal model with autoregressive decoding. This structure makes inference memory-bandwidth-bound and pushes per-decision latency into a regime where comfortable onboard execution is no longer obvious even on the most capable automotive SoCs. The gain is a richer driving model that can reason and explain, but the deployment burden grows with required processor size and memory bandwidth, which is what makes a vehicle-versus-cloud split worth analyzing. However, most autonomous-driving literature assumes the model runs onboard and characterizes inference cost on a single GPU; it leaves open whether a fleet should divide the pipeline between vehicle and cloud and how that division changes bandwidth, latency, and the leftover in-vehicle hardware.

\subsection{Communication infrastructure.} This literature sets the network side of the problem by tracing how each generation of vehicular wireless can support the uplink that cloud driving requires. The earliest tier, such as dedicated short-range communications (DSRC)~\citep{naik2019ieee80211bd} and cellular vehicle-to-everything (C-V2X)~\citep{kumar2023cv2x}, provides reliable low-latency channels intended for short safety messages and cooperative awareness, but its bandwidth is small relative to the data rates AV inference offloading would require. A second tier introduces new radio vehicle-to-everything (NR-V2X) and 5G ultra-reliable low-latency communication (URLLC) under 3GPP Release~16+~\citep{3gpp2021ts38300, garcia2021tutorial}, which combine sub-millisecond air-interface access with higher-throughput uplink and begin to make continuous high-rate data streams plausible. 5G-Advanced under Release~18/19~\citep{lin2022overview5gadv} builds on this with enhanced uplink scheduling, multi-cell coordination, and tighter URLLC guarantees, narrowing the gap between cellular capability and the latency budgets a closed driving loop requires. Most recently, 6G vehicular intelligence proposals~\citep{itu2023imt2030, saad2020vision6g, dang2020what6g, tataria20216gwireless, guo2022vehicular6g} target sub-100~microsecond air-interface delay, terabit-per-second peak rates, and AI-native scheduling, framing the vehicular link itself as part of a computational fabric rather than a passive pipe. Across generations, these studies clarify scheduling, reliability, multiple input multiple output (MIMO), and edge-native networking at the link level. However, their traffic examples are usually short safety messages, cooperative awareness packets, or limited perception sharing rather than a continuous high-rate uplink from AVs to cloud inference. Measurements from 5G vehicular testbeds~\citep{santa2022evaluation} and mobile edge computing surveys~\citep{mao2017surveymec, liu2019edgedriving} show that edge hosting can reduce delay, but the deployment question is sharper: how many active vehicles in the same cell can each sustain the uplink rate required by a chosen offloading strategy while still meeting the closed-loop deadline? Section~\ref{sec:comm_model} formalizes this conversion from link performance to per-cell admission capacity, which matters because cloud driving turns the usual cellular pattern around---the bottleneck is the uplink from vehicle to infrastructure rather than the downlink to the user.

\subsection{Vehicular edge computing and task offloading.} This literature asks where vehicle-related computation should run when the link, the edge, and the cloud each have different capacity and cost profiles. Several mechanism families have emerged. Mobility-aware schedulers~\citep{li2024mobility} place tasks at edge servers that anticipate the vehicle's path, reducing handover-induced delay when a vehicle moves between cells during a single offloaded task. Multi-path offloading~\citep{liu2024multipath} splits a single task across heterogeneous links so that aggregate bandwidth grows and transient outages on one link are masked by the others. Cost-aware Stackelberg formulations~\citep{wang2024costaware} model the operator--vehicle interaction as a leader--follower pricing game, recovering decentralized equilibria when each agent optimizes its own delay--cost tradeoff. Network slicing strategies~\citep{hejja2022slicing} reserve dedicated radio and transport resources for vehicular offloading traffic so that flows are isolated from other tenants on the same infrastructure. Deep reinforcement learning controllers~\citep{mekrache2022drl} learn offloading policies directly from interaction with the environment, adapting to non-stationary congestion patterns that closed-form models cannot capture. Together, these works show that the offloading decision depends jointly on congestion, delay, and price. These are the closest relatives of this paper, but they often describe the workload as one generic task, which hides the important differences among E2E, VLM, and VLA models. Those model classes differ in compute structure, pressure on GPU memory bandwidth, autoregressive decoding patterns, and the split point where part of the driving pipeline can leave the vehicle---differences that change both what bandwidth is needed on the uplink and what service rate is needed at the GPU. The economic version of the question is also fleet-level rather than task-level: should every vehicle carry expensive peak hardware, or should the operator buy a shared GPU pool whose queueing tail---the rare but safety-relevant waiting time at the end of the delay distribution---still fits the driving latency budget?

\subsection{Synthesis.} Our framework treats these three literatures as inputs to one deployment sequence: the driving-model split sets the workload, the communication system sets whether that workload can be uploaded under a closed-loop deadline, and the cloud GPU system sets whether shared service can absorb the resulting demand. Section~\ref{sec:framework} formalizes this sequence; Section~\ref{sec:case_studies} applies it to the New York City case study introduced in Section~\ref{sec:introduction}.

\begin{figure}[width=0.99\textwidth, pos = t]
\centering
\includegraphics[width=0.99\textwidth]{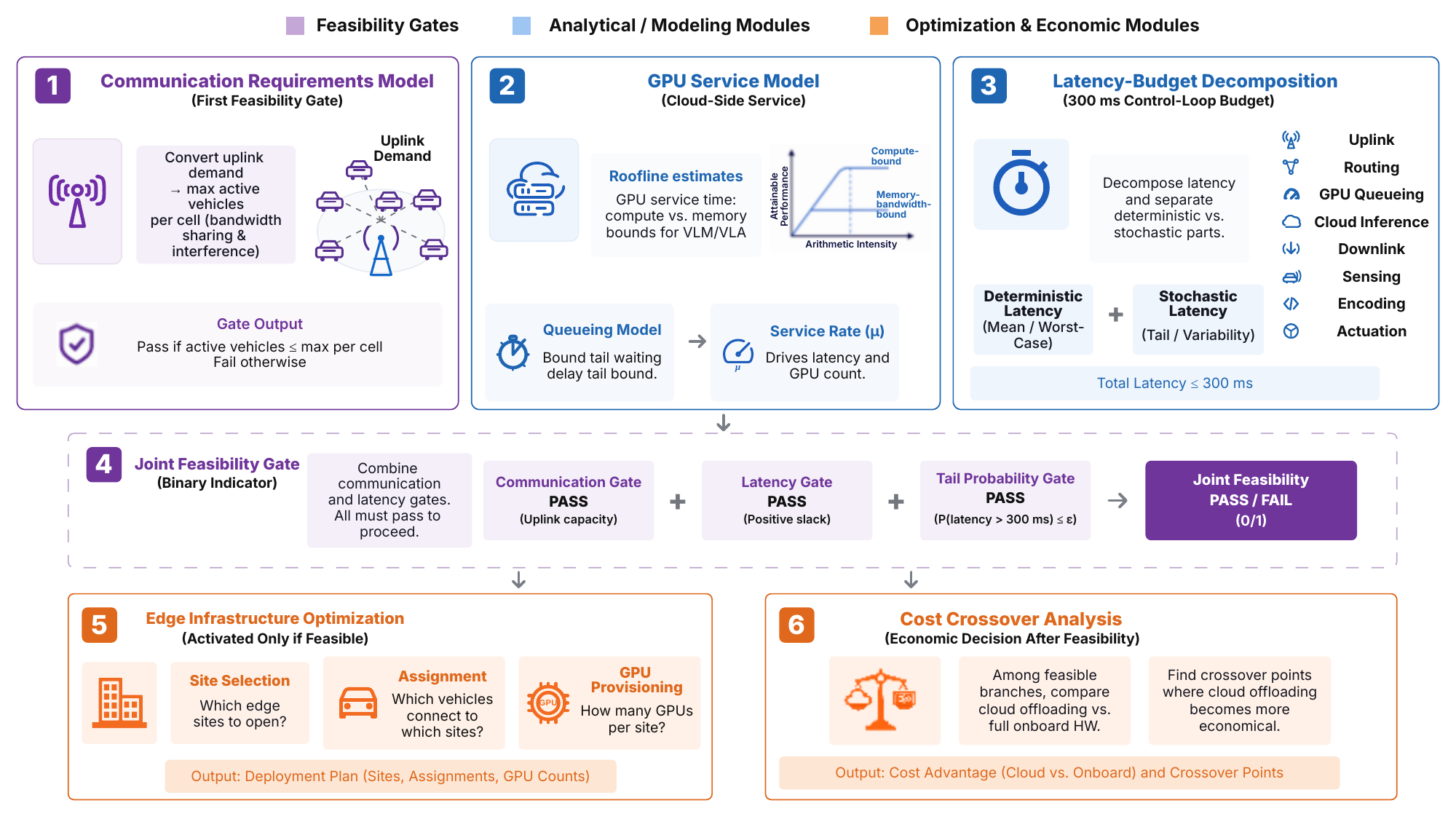}
\caption{Analytical pipeline. The Communication Requirements Model defines the communication-side limit, and the GPU Service Model defines the cloud-side service model used by the Stochastic Latency Model. Those results then feed the Feasibility Threshold; feasible branches proceed to edge infrastructure optimization and are finally compared through the Total Cost of Ownership and Crossover step.}
\label{fig:flowchart}
\end{figure}

\section{Analytical Framework}
\label{sec:framework}

Figure~\ref{fig:flowchart} summarizes the analytical framework, moving from communication, GPU service model, latency, and feasibility to infrastructure optimization and cost analysis. Section~\ref{sec:scenario_inputs} begins this sequence by specifying the scenario inputs: AV penetration rate, utilization, communication generation, offloading strategy, and model class.
These inputs parameterize each evaluated scenario.
Section~\ref{sec:offloading} then maps each offloading strategy to cloud-side and onboard inference portions, identifying the residual onboard hardware required by the onboard portion. 
Section~\ref{sec:comm_model} next evaluates communication generation and access-network admission for the cloud-side inference portion. Section~\ref{sec:gpu_service_model} defines the GPU Service Model, which converts cloud-side inference demand into service rate, inference latency, and queueing latency under tail-provisioning. The Stochastic Latency Model in Section~\ref{sec:stochastic_latency} and Feasibility Threshold in Section~\ref{sec:feasibility} then combine communication delay and GPU-service latency into a single feasibility gate.
Feasible scenarios then proceed to edge infrastructure optimization (EIO) in Section~\ref{sec:edge_placement}, Total Cost of Ownership and Crossover in Section~\ref{sec:tco}, and the three binding-regime New York City case study in Section~\ref{sec:case_studies}.

\subsection{Scenario Space and Inputs}
\label{sec:scenario_inputs}

In order to optimize edge infrastructure placement, we need three inputs: communication generation (Section~\ref{sec:comm_generation_inputs}), fleet demand (Section~\ref{sec:fleet_demand_parameters}), and model class (Section~\ref{sec:model_class_parameters}). Communication generation input defines parameters for different communication characteristics. Fleet demand combines AV penetration rate and utilization level to determine expected cell loading. The model class distinguishes E2E, VLM, and VLA to capture workloads with different compute needs.

\subsubsection{Communication Generation Inputs}
\label{sec:comm_generation_inputs}
Each input for communication generation $g\in\{5\mathrm{G},5\mathrm{G\mbox{-}Adv},6\mathrm{G}\}$ specifies bandwidth, user-plane latency, reliability, connection density, and uplink-tail assumptions for cloud inference uploads.
 
Together, these assumptions determine the uplink headroom and delay tails per-cell. Thus, each communication-generation case carries the bandwidth, latency, reliability, density, spectrum, and uplink-tail assumptions used
downstream. Holding these communication-generation assumptions fixed isolates how fleet demand, offloading strategy, and model class affect feasibility and cost between scenarios. The communication-generation parameters are summarized in Table~\ref{tab:comm_gen}. These parameters will be used in Section \ref{sec:comm_model}.

\begin{table}[h]
\caption{This table lists
the corresponding bandwidth, latency, reliability, density, spectrum, and uplink-tail values for each communication generation. 5G values reflect deployed FR2/URLLC capabilities~\citep{3gpp2021ts38300,garcia2021tutorial}; 5G-Advanced reflects 3GPP Release 18/19 targets~\citep{lin2022overview5gadv}; 6G reflects ITU IMT-2030 aspirational targets~\citep{itu2023imt2030}.}
\label{tab:comm_gen}
\centering
\small
\begin{tabular}{@{}lccc@{}}
\toprule
\textbf{Parameter} & \textbf{5G (FR2)} & \textbf{5G-Adv (R18/19)} & \textbf{6G (IMT-2030)} \\
\midrule
Peak downlink & 20~Gbps & 30+~Gbps & 1~Tbps \\
Peak uplink & 10~Gbps & 15+~Gbps & 200+~Gbps \\
User-exp.\ uplink & 10--50~Mbps & 50--200~Mbps & 1+~Gbps \\
Latency (user plane) & 1--4~ms & $<$1~ms & 0.1~ms \\
Reliability & 99.999\% & 99.9999\% & 99.99999\% \\
Connection density & $10^6$/km$^2$ & $10^6$/km$^2$ & $10^7$/km$^2$ \\
US spectrum (mmWave) & 24--47~GHz & + 7--24~GHz (FR3) & + sub-THz \\
\midrule
$\lambda_{\text{UL}}^{(g)}$ (s$^{-1}$) & $\approx$500 & $\approx$2{,}000 & $\approx$10{,}000 \\
\bottomrule
\end{tabular}
\end{table}

\subsubsection{Fleet Demand Parameters}
\label{sec:fleet_demand_parameters}
We parametrize fleet demand using AV penetration $\rho$ and utilization $u$ to compute active vehicle counts. AV penetration takes $\rho \in \{0.1\%, 1\%, 5\%, 10\%, 20\%, 30\%, 50\%, 100\%\}$ and utilization takes $u \in \{0.05, 0.12, 0.30, 0.45, 0.65, 1.0\}$. The $u=0.05$ case anchors the personal-vehicle baseline, reflecting the fact that personal vehicles are parked 95\% of the time~\citep{doe2024fotw1356}.
The $u=0.12$ case represents a conservative upper-end personal-use scenario.
Moving up the utilization sweep, $u=0.30$ and $u=0.45$ capture mixed-fleet operation~\citep{karolemeas2024shared_u_30_45}, while $u=0.65$ captures robotaxi-level utilization~\citep{yang_u_65}.
 The $u=1.0$ case represents full use, with every vehicle active 100\% of the time. This sweep maps each $(\rho,u)$ scenario to one active vehicle count.
 That active vehicle count is then used consistently across the three binding regimes of the New York City case study.

\subsubsection{Model Class Parameters}
\label{sec:model_class_parameters}
The model class $m\in\{\mathrm{E2E},\mathrm{VLM},\mathrm{VLA}\}$ selects the cloud inference workload and onboard baseline values. For E2E, VLM, and VLA, these values set cloud FLOPs and inference time for feasibility analysis, while also setting full onboard hardware cost for cost crossover. 

\subsection{Offloading Strategy Spectrum}
\label{sec:offloading}

\begin{figure}[width=0.99\textwidth, pos = t]
    \centering
    \resizebox{0.7\columnwidth}{!}{

\definecolor{vehblue}{HTML}{0072B2}      
\definecolor{cloudorange}{HTML}{E69F00}   
\definecolor{uplinkred}{HTML}{D55E00}     
\definecolor{downlinkgreen}{HTML}{009E73}  
\definecolor{lightgray}{HTML}{F0F0F0}

\begin{tikzpicture}[
    >=Stealth,
    pblock/.style={
        rectangle, rounded corners=2pt, minimum height=0.72cm,
        font=\scriptsize\sffamily, align=center, line width=0.5pt,
        inner xsep=4pt, inner ysep=2pt,
    },
    veh/.style={pblock, draw=vehblue!70, fill=vehblue!15, minimum width=1.5cm},
    cld/.style={pblock, draw=cloudorange!70, fill=cloudorange!15,minimum width=1.5cm},
    slabel/.style={font=\small\sffamily\bfseries, anchor=east},
    annot/.style={font=\tiny\sffamily, text=black!60, align=center},
    uplink/.style={->, thick, color=uplinkred, line width=1pt},
    downlink/.style={->, thick, color=downlinkgreen, line width=1pt},
    cutline/.style={densely dashed, line width=0.8pt, color=black!40},
]

\def\colA{0}      
\def\colB{1.9}    
\def\colC{3.8}    
\def\colD{5.7}    
\def\colE{7.6}    

\def\rowS{0}       
\def\rowSS{-1.6}   
\def\rowSSS{-3.2}  
\def\rowSSSS{-4.8} 

\node[font=\scriptsize\sffamily\bfseries, text=black!70, align=center] at (\colA, 1.0) {Sensors};
\node[font=\scriptsize\sffamily\bfseries, text=black!70, align=center] at (\colB, 1.0) {Vision\\Backbone};
\node[font=\scriptsize\sffamily\bfseries, text=black!70, align=center] at (\colC, 1.0) {Transformer\\Encoder};
\node[font=\scriptsize\sffamily\bfseries, text=black!70, align=center] at (\colD, 1.0) {Decoder /\\Planning};
\node[font=\scriptsize\sffamily\bfseries, text=black!70, align=center] at (\colE, 1.0) {Vehicle\\Control};

\draw[black!20, line width=0.4pt] (-0.9, 0.55) -- (8.6, 0.55);

\node[slabel] at (-1.1, \rowS) {\textsf{S1}};

\node[veh] (s1a) at (\colA, \rowS) {Sensor\\capture};
\node[cld] (s1b) at (\colB, \rowS) {Vision\\backbone};
\node[cld] (s1c) at (\colC, \rowS) {Transformer\\encoder};
\node[cld] (s1d) at (\colD, \rowS) {Decoder /\\planning};
\node[veh] (s1e) at (\colE, \rowS) {Vehicle\\control};

\draw[uplink] (s1a.east) -- (s1b.west)
    ;
\draw[->, thick, color=cloudorange!50] (s1b.east) -- (s1c.west);
\draw[->, thick, color=cloudorange!50] (s1c.east) -- (s1d.west);
\draw[downlink] (s1d.east) -- (s1e.west);


\node[slabel] at (-1.1, \rowSS) {\textsf{S2}};

\node[veh] (s2a) at (\colA, \rowSS) {Sensor\\capture};
\node[veh] (s2b) at (\colB, \rowSS) {Vision\\backbone};
\node[cld] (s2c) at (\colC, \rowSS) {Transformer\\encoder};
\node[cld] (s2d) at (\colD, \rowSS) {Decoder /\\planning};
\node[veh] (s2e) at (\colE, \rowSS) {Vehicle\\control};

\draw[->, thick, color=vehblue!50] (s2a.east) -- (s2b.west);
\draw[uplink] (s2b.east) -- (s2c.west)
    ;
\draw[->, thick, color=cloudorange!50] (s2c.east) -- (s2d.west);
\draw[downlink] (s2d.east) -- (s2e.west);



\node[slabel] at (-1.1, \rowSSS) {\textsf{S3}};

\node[veh] (s3a) at (\colA, \rowSSS) {Sensor\\capture};
\node[veh] (s3b) at (\colB, \rowSSS) {Vision\\backbone};
\node[veh] (s3c) at (\colC, \rowSSS) {Transformer\\encoder};
\node[cld] (s3d) at (\colD, \rowSSS) {Decoder /\\planning};
\node[veh] (s3e) at (\colE, \rowSSS) {Vehicle\\control};

\draw[->, thick, color=vehblue!50] (s3a.east) -- (s3b.west);
\draw[->, thick, color=vehblue!50] (s3b.east) -- (s3c.west);
\draw[uplink] (s3c.east) -- (s3d.west)
    ;
\draw[downlink] (s3d.east) -- (s3e.west);


\draw[black!30, line width=0.6pt, densely dashed]
    (-0.9, {0.5*(\rowSSS+\rowSSSS)}) -- (8.6, {0.5*(\rowSSS+\rowSSSS)});

\node[slabel, text=black!60] at (-1.1, \rowSSSS) {\textsf{Baseline}};

\node[veh] (s4a) at (\colA, \rowSSSS) {Sensor\\capture};
\node[veh] (s4b) at (\colB, \rowSSSS) {Vision\\backbone};
\node[veh] (s4c) at (\colC, \rowSSSS) {Transformer\\encoder};
\node[veh] (s4d) at (\colD, \rowSSSS) {Decoder /\\planning};
\node[veh] (s4e) at (\colE, \rowSSSS) {Vehicle\\control};

\draw[->, thick, color=vehblue!50] (s4a.east) -- (s4b.west);
\draw[->, thick, color=vehblue!50] (s4b.east) -- (s4c.west);
\draw[->, thick, color=vehblue!50] (s4c.east) -- (s4d.west);
\draw[->, thick, color=vehblue!50] (s4d.east) -- (s4e.west);


\draw[<->, line width=1.2pt, color=black!50]
    (-1.8, \rowS+0.4) -- (-1.8, \rowSSS-0.4);
\node[font=\tiny\sffamily, text=cloudorange!80!black, rotate=90, anchor=south]
    at (-2.15, \rowS-0.3) {More cloud};
\node[font=\tiny\sffamily, text=vehblue!80!black, rotate=90, anchor=north]
    at (-2.55, \rowSSS+0.3) {More vehicle};

\node[veh, minimum width=1.2cm, minimum height=0.4cm] (legv) at (0.0, -6.0) {};
\node[font=\tiny\sffamily, right=0.1cm of legv] {In-vehicle};
\node[cld, minimum width=1.2cm, minimum height=0.4cm] (legc) at (2.5, -6.0) {};
\node[font=\tiny\sffamily, right=0.1cm of legc] {Cloud / edge};

\draw[uplink] (4.5, -6.0) -- ++(0.7,0);
\node[font=\tiny\sffamily, anchor=west] at (5.3, -6.0) {Uplink};

\draw[downlink] (6.5, -6.0) -- ++(0.7,0);
\node[font=\tiny\sffamily, anchor=west] at (7.3, -6.0) {Downlink};

\end{tikzpicture}
    }
\caption{Offloading strategy spectrum.}
\label{fig:offloading_spectrum}
\end{figure}
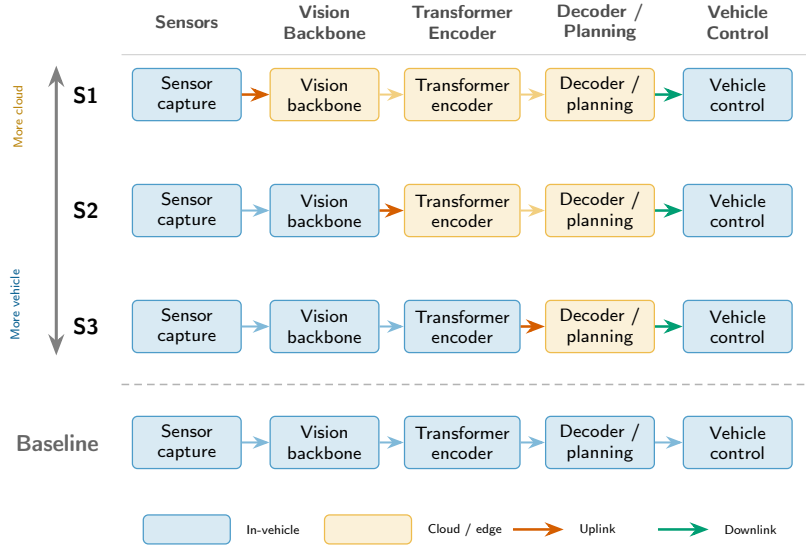

We begin by specifying the offloading strategy, which maps each AD pipeline stage to onboard execution or cloud inference. To ground this mapping, Figure~\ref{fig:offloading_spectrum} orders the AD pipeline stages from sensor capture through vehicle control. Within this ordered pipeline, \textit{S1}/\textit{S2}/\textit{S3} offload raw sensor data, compressed features, and scene queries, respectively. These offload points specify what leaves the vehicle, what remains onboard, and the residual in-vehicle compute required for communication, latency, and cost calculations.

%

Figure~\ref{fig:offloading_spectrum} localizes S1/S2/S3 within the AD pipeline, while Table~\ref{tab:offloading_spectrum} specifies the uplink, in-vehicle compute, and cloud-side workload assumptions.
 We report in-vehicle compute in TOPS (INT8) and cloud compute in TFLOPS (FP16), treating them as separate inputs to feasibility and cost calculations.
 This separation makes the $S1$--$S3$ tradeoff explicit: lower uplink demand requires more stages to remain onboard, which increases residual in-vehicle compute and limits hardware-cost savings from cloud offloading. We track this residual in-vehicle compute in TOPS and include
it in the total cost of ownership (TCO) analysis in Section~\ref{sec:tco}. Table~\ref{tab:offloading_spectrum} then summarizes how residual in-vehicle TOPS, uplink demand,
and cloud workload parameterize the offloading strategy spectrum in the experiments.


Model diversity within each class motivates using representative references for the analytical framework. We estimate TOPs using UniAD~\citep{hu2023uniad}, DriveLM~\citep{sima2024drivelm}, and Alpamayo~\citep{nvidia2026alpamayo} as references.
 We convert $\mathrm{TFLOPs}$ to $\mathrm{TOPs}$ using the 2:1 ratio between INT8 and FP16 throughput reported on current GPU datasheets~\citep{nvidia2025rtxpro6000}. Table~\ref{tab:model_component_sizes} summarizes TFLOPs for each phase of the models, so alternative models can enter the same analytical framework. Alternative models require only updating Table~\ref{tab:model_component_sizes}, not the analytical framework.

\begin{table}[width=\textwidth, pos=b]
\centering
\caption{Compute requirements (in TFLOPs) for each phase across different models.}
\label{tab:model_component_sizes}
\begin{tabular}{lccc}
\toprule
\textbf{Architecture} &
\textbf{Vision Backbone} &
\textbf{Transformer Encoder} &
\textbf{Decoder / Planning} \\
\midrule
E2E & 0.31  & 1.33   & 0.06  \\
VLM & 4.53 & 19.35  & 0.82 \\
VLA & 11.00 & 47.00 & 2.00 \\
\bottomrule
\end{tabular}
\end{table}

\begin{table}[width=\textwidth, pos=b]
\caption{The table lists each strategy with its uplink range,
experimental uplink value, residual INT8 TOPS, and FP16 cloud FLOPs for E2E, VLM, and VLA models. The uplink range reflects typical sensor and feature payload sizes at $f = 10$~Hz planning frequency (6 cameras at 1080p/30fps, 200K-point LiDAR stream); the experiments use the single value in the ``Used'' column. Residual in-vehicle compute $H_s$ and cloud FLOPs depend on the model class; values are listed as E2E / VLM / VLA. Note that while $H_s$ accounts for frequency, cloud FLOPs are measured per inference since cloud GPUs process requests independently.}
\label{tab:offloading_spectrum}
\centering
\footnotesize
\begin{tabular}{@{}l c c c c@{}}
\toprule
\textbf{Strategy} & \textbf{Uplink range} & \textbf{Used} & \textbf{Residual $H_s$} & \textbf{Cloud FLOPs} \\
                  & \textbf{(Mbps)}       & \textbf{(Mbps)} & \textbf{(TOPS, INT8)} & \textbf{(TFLOPs, FP16)} \\
\midrule
S1 (Raw Sensor)     & 50--200 & 100 & 5 / 5 / 5          & 1.7 / 24.7 / 60.0 \\
S2 (Feature-Level)  & 10--25  & 25  & 16 / 226 / 550       & 1.39 / 20.17 / 49.0 \\
S3 (Query-Level)    & 1--5    & 3   & 82 / 1194 / 2900      & 0.06 / 0.82 / 2.0 \\
\midrule
In-vehicle baseline & ---     & --- & 85 / 1235 / 3{,}000 & --- \\
\bottomrule
\end{tabular}
\end{table}

\subsubsection{Raw-Sensor Offloading}
\label{sec:Raw-Sensor_Offloading}
Strategy S1 (Raw Sensor Offloading) offloads raw multimodal sensor data directly to the cloud for cloud inference. Because S1 moves the pre-perception input rather than an intermediate representation, we size its communication generation load using 6 cameras at 1080p/30fps (H.265, 30--120~Mbps), a 200K-point LiDAR stream (20--80~Mbps compressed), and radar ($\sim$0.5~Mbps), yielding a total uplink of \textbf{50--200~Mbps} at a 10~Hz planning frequency; the experiments use $B_{S1} = 100$~Mbps where $B_s$ is the strategy-specific target uplink rate. This high uplink requirement shows that S1 primarily trades onboard perception, prediction, and planning compute for communication demand. This split keeps only encoding onboard ($H_{S1} = 5$~TOPS across all model classes), making communication generation the central constraint. 

\subsubsection{Feature-Level Offloading}
\label{sec:Feature_Level_Offloading}

Strategy S2 (Feature-Level Offloading) keeps the vision backbone local and uploads compressed intermediate features at \textbf{10--25~Mbps}, 4--8$\times$ below S1. In this strategy, the vision backbone refers to the early perception stages that convert raw sensor input into intermediate scene features before offloading. Keeping the vision backbone onboard limits residual demand to $H_{S2} = 16$, $226$, and $550$~TOPS for E2E, VLM, and VLA, respectively. With the backbone kept
local, S2 offloads the E2E planning head or the VLM/VLA language decoder, which accounts for most remaining FLOPs. The experiments use $B_s = 25$~Mbps. 

\subsubsection{Query-Level Offloading}
\label{sec:Query_Level_Offloading}
Strategy S3 (Query-Level Offloading) runs the transformer encoder onboard, then uploads compact scene query tokens at \textbf{1--5~Mbps} (the experiments use $B_{S3} = 3$~Mbps). This split makes the uploaded scene query tokens compact while retaining
substantial encoder compute onboard. The compact uplink makes S3 easy for the communication system to support under scarce connectivity, so its main limitation shifts from access-network admission to residual vehicle compute.
 Residual vehicle compute remains high: $H_{S3}=82$, $1194$, and $2900$~TOPS for E2E, VLM, and VLA.

\subsubsection{In-Vehicle Baseline}
\label{sec:In-Vehicle_Baseline}
We denote the peak full-pipeline onboard requirements for model class $m$ by $H_m^{\text{full}}$, allowing later cost equations to price the hardware each vehicle must carry for full in-vehicle inference.
For this baseline, full-pipeline onboard requirements are $H_{\text{E2E}}^{\text{full}} = 85$~TOPS, $H_{\text{VLM}}^{\text{full}} = 1235$~TOPS, and $H_{\text{VLA}}^{\text{full}} = 3{,}000$~TOPS.
 This notation defines the full in-vehicle inference baseline, whose
independence from networks makes it the reference case for safety reasoning. 

\subsection{Communication Requirements Model}
\label{sec:comm_model}

Each S1--S3 strategy is evaluated using two access-network quantities. The first is the \textit{per-vehicle bandwidth requirement} implied by each strategy. The second is the \textit{cell-level capacity limit}: the active-vehicle load one access cell can admit under that per-vehicle bandwidth requirement. Accordingly, the Communication Requirements Model links uplink demand to access-network admission. Applying this feasibility setup to the three strategies yields sustained per-vehicle uplink rates of \textbf{100~Mbps} for S1, \textbf{25~Mbps} for S2, and \textbf{3~Mbps} for S3; downlink remains 1~Mbps. Because the 1~Mbps downlink is small relative to these
sustained uplink rates, admission is driven primarily by uplink demand. Accordingly, this section also maps sustained per-vehicle uplink demand for each offloading option to cell capacity through spectral efficiency rather than bandwidth alone. As a reference point for this mapping, a common baseline fixes spectral efficiency per cell~\citep{garcia2021tutorial}, making capacity proportional to bandwidth. However, simultaneous vehicle uploads make spectral efficiency load-dependent as resource sharing and interference reduce bits per second per hertz. We model this with interference-aware orthogonal frequency-division multiple access (OFDMA) uplink~\citep{andrews2011tractable, 3gpp2020tr38901}.

\subsubsection{Uplink Demand Model}
At the cell level, scheduling determines per-vehicle uplink allocation. We use proportional-fair scheduling as the allocation rule, converting generation-specific uplink capacity and channel quality into the communication gate's per-vehicle capacity input. For a cell with $n$ co-scheduled vehicles, this scheduling rule determines the per-vehicle allocation used to test whether each offloading option sustains its uplink demand. Formally, the per-vehicle bandwidth allocation is:
\begin{equation}
\label{eq:resource_alloc}
W_v(n, g) = W_c^{(g)} / n,
\end{equation}
where $n$ is the number of vehicles sharing the cell, $g$ is the communication generation, and $W_c$ is the cell bandwidth.

\subsubsection{Interference-Aware Cell Capacity}

To capture the impact of multi-user interference, we model the cell-average uplink signal-to-interference-plus-noise ratio (SINR) after beamforming for active vehicles with $M^{(g)}$ gNB antenna elements as
\begin{equation}
\label{eq:sinr}
\bar{\gamma}(n, g) = \frac{\gamma_0^{(g)}}{1 + (n{-}1)\,\alpha^{(g)} / M^{(g)}},
\end{equation}
where $n$ is the number of vehicles sharing the cell, $g$ is the communication generation, $\gamma_0^{(g)}$ is the 3GPP UMa single-user SINR, and $\alpha^{(g)}$ is the intra-cell interference after MIMO spatial separation.

Equation~\eqref{eq:sinr} combines 3GPP UMa single-user SINR with residual intra-cell interference after MIMO spatial separation~\citep{3gpp2020tr38901,andrews2011tractable}. This expression therefore augments the 3GPP UMa single-user SINR with a residual-interference penalty, $(n{-}1)\alpha/M$, that scales with co-scheduled vehicles. This effective SINR then enters the implementation-adjusted spectral-efficiency calculation before per-user bandwidth sharing is applied. Under uniform power allocation across $K^{(g)}$ spatial streams, this calculation gives 
\begin{equation}
\label{eq:spec_eff}
\eta_v(n,g)=K^{(g)}\log_2\!\left(1+\frac{\bar{\gamma}(n,g)}{K^{(g)}\Gamma^{(g)}}\right), 
\end{equation}
and bandwidth sharing yields 
\begin{equation}
\label{eq:per_user_bandwidth}
R_v(n,g)=\frac{W_c^{(g)}}{n}\eta_v(n,g).
\end{equation}
Accordingly, the communication model compares each strategy’s uplink demand with the practical per-user rate, $R_v(n,g)$, rather than an ideal link rate, because that rate already incorporates implementation loss, spatial-stream sharing, and bandwidth sharing. Thus, the practical per-user rate, $R_v(n,g)$, couples bandwidth sharing with SINR-dependent spectral efficiency, making cell capacity load-dependent. This dependence links the achievable rate directly to the strategy-specific target rate $B_s$, motivating an admission threshold indexed by $s$ and $g$. 
\begin{equation}
\label{eq:cell_capacity}
N_c^{\max}(s, g) = \max\!\left\{ n \in \mathbb{Z}_+ : R_v(n, g) \geq B_s \right\}.
\end{equation}
 
\begin{figure}[width=0.99\textwidth, pos = t]
\centering
\includegraphics[width=0.50\linewidth]{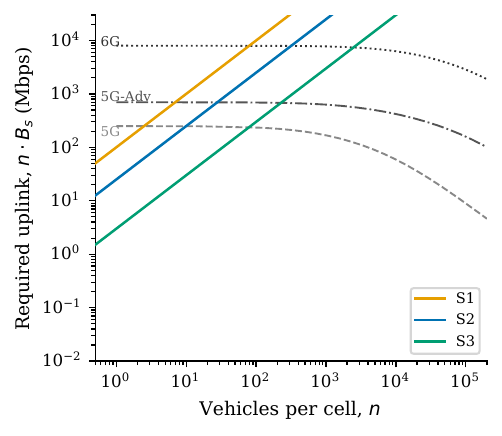}
\caption{Bandwidth-capacity tradeoff. Colored lines show the required uplink demand for S1--S3 as vehicles per cell increase, while gray generation curves show the available uplink capacity under the interference-aware cell model. Intersections mark the maximum bandwidth-feasible vehicles per cell.}
\label{fig:bandwidth_capacity}
\end{figure}

 Equivalently, Eq.~\eqref{eq:cell_capacity} defines this admission threshold as an integer feasibility problem: among candidate vehicle counts $n$, select the largest count for which the interference-aware per-user rate meets the strategy-specific target rate, $R_v(n,g)\ge B_s$. Since $R_v(n,g)$ declines monotonically in $n$, integer bisection finds this largest feasible count. The resulting value defines $N_c^{\max}(s,g)$; Figure~\ref{fig:bandwidth_capacity} shows that, for 5G-Advanced S2 ($B_s = 25$~Mbps), interference reduces the feasible count to $\sim$27, a $\sim$4\% tighter bound.

\subsection{GPU Service Model}
\label{sec:gpu_service_model}

The GPU Service Model defines queueing latency and inference latency for the offloaded workload.
 It derives these latencies from the GPU Queueing Model (Section~\ref{sec:GPU_Queueing_Model}) and the Roofline Inference Model (Section~\ref{sec:Roofline_Inference_Model}).
Together,these derived latencies provide the cloud-
side values used for the feasibility analysis in later sections

\subsubsection{GPU Queueing Model}
\label{sec:GPU_Queueing_Model}
Each edge site $e$ is an M/D/$c_e$ queue with aggregate arrival rate:
\begin{equation}
\label{eq:arrival_rate}
\Lambda_e=f\sum_{v\in\mathcal{V}} y_{ve},
\end{equation}
where $\mathcal{V}$ is the candidate vehicle set, $y_{ve}$ is the vehicle-to-site assignment indicator and $f$~[Hz] is the planning frequency.

In other words, $\Lambda_e$ is the load from vehicles assigned to site $e$. We define the per-server load as
\begin{equation}
\rho_e=\frac{\Lambda_e}{c_e\mu_{\text{eff}}(s,m,t)},
\end{equation}
where $c_e$ is the number of GPUs at site $e$. Each GPU has deterministic service rate $\mu_{\text{eff}}(s,m,t)$ from the roofline model in Section~\ref{sec:Roofline_Inference_Model}.

Formally, the required GPU count is defined as the smallest \(c_e\) satisfying the queueing-tail constraint, 
\begin{equation}
\label{eq:min_gpus}
c_e^*(s,m,t) = \min\!\left\{ c_e \in \mathbb{Z}_+ : P_Q(c_e, \rho_e) \cdot e^{-c_e \mu_{\text{eff}}(s,m,t) (1 - \rho_e) L_{\text{q}}^{\max}} \leq \epsilon_q \right\},
\end{equation}
where $P_Q$ is the Erlang-C queueing probability, $\epsilon_q$ is the tail-delay bound, and $L_{\text{q}}^{\max}$ is the maximum affordable queueing delay. $c_e^*(s,m,t)$ ensures $\rho_e<1$. More precisely, Eq.~\eqref{eq:min_gpus} chooses the smallest \(c_e\) that keeps the waiting-time tail at site $e$ below the target bound, using the Erlang-C waiting probability and the exponential waiting-duration bound. Specifically, \(P_Q\) gives the Erlang-C probability that a planning request waits, and the exponential factor gives the conditional tail probability that the wait exceeds \(L_{\text{q}}^{\max}\).
Since increasing \(c_e\) lowers the Erlang-C waiting probability and the conditional tail term, we compute the minimum feasible GPU count by bisection before evaluating its overhead relative to mean-load sizing. 
At the target tail-delay bound ($\epsilon_q=10^{-5}$), this bisection-based sizing requires on the order of 15--25\% more GPUs than mean-load sizing at typical multi-site loading, dropping to a few percent at the most heavily loaded sites and rising for lightly loaded sites.

\subsubsection{Roofline Inference Model}
\label{sec:Roofline_Inference_Model}
Autoregressive VLM/VLA inference combines GPU arithmetic work with repeated high-bandwidth memory (HBM) reads. We model its service time as the sum of compute-bound and memory-bandwidth-bound terms.
This decomposition gives the effective service time for offloading strategy $s$ and model class $m$:
\begin{equation}
\label{eq:roofline}
\mu_{\text{eff}}^{-1}(s,m,t) = \underbrace{\frac{C_{\text{encoder}}(s,m) + C_{\text{prefill}}(s,m)}{\phi_{\text{GPU}}(t)}}_{\text{compute-bound}} + \underbrace{\frac{D(s,m) \cdot W_{\text{reasoning}}(s,m) + n_{\text{traj}}(s,m) \cdot W_{\text{traj}}(s,m)}{B_{\text{mem}}(t)}}_{\text{bandwidth-bound}},
\end{equation}
where $ C_\text{encoder} $ and $C_\text{prefill}$ are FLOP counts of the encoding and prefill stages. $D$ is the number of reasoning tokens, $W_\text{reasoning}$ and $W_\text{traj}$ are the reasoning and trajectory weight footprints per step, respectively, and $n_\text{traj}$ is the number of trajectory rollout steps. $\phi_{\text{GPU}}(t)$ and $B_{\text{mem}}$ are the GPU compute--throughput and HBM-bandwidth at a given time $t$, respectively, which are discussed in Section~\ref{sec:gpuevol}.

In this equation, the GPU compute-throughput and HBM-bandwidth parameters translate model-specific workloads into compute time and HBM-read time at time $t$. Our speed profiling for Alpamayo~\citep{nvidia2026alpamayo} justifies the importance of the bandwidth term in this equation: reasoning and trajectory decoding take roughly 80\% of wall-clock inference time but only $\sim$3\% of forward-pass FLOPs. By contrast, for a non-autoregressive model ($D=0$, $n_{\text{traj}}=0$), the memory-bound terms vanish, so the service time reduces to compute throughput divided by offloaded FLOPs.

To show why this compute-only approximation fails for autoregressive VLAs, applying the B300 (2025) baseline compute term to Alpamayo yields a
latency of $39$~ms. However, the revised decomposition also includes the memory-bound decoding workload, adding the corresponding per-step token counts and HBM-read volumes ($D=16$ reasoning tokens, $W_{\text{reasoning}}=15.6$~GB, $n_{\text{traj}}=17$ trajectory steps, and $W_{\text{traj}}=22.9$~GB per step). The revised decomposition yields a B300 inference latency of $153$~ms (raw-sensor offloading; $\sim$114~ms of it is bandwidth-bound decode). The $153$~ms latency corresponds to $6.5$ inferences per second, in line with measured driving-model inference rates such as $1.8$~FPS for UniAD-Single and $0.16$~FPS for the autoregressive DriveLM-Agent~\citep{sima2024drivelm}.

\subsubsection{GPU Evolution Model}
\label{sec:gpuevol}

To project both compute and memory bottlenecks, we let GPU performance follow the historical A100$\to$H100$\to$B300 trend and project forward from the B300 baseline year $t_0=2025$, while annual growth rates decelerate as semiconductor scaling matures. We model the GPU evolution rate at time $t$ by:
\begin{equation}
\label{eq:gpu_evolution_rate}
r_\phi(t) = r_\phi^\infty + (r_\phi^0 - r_\phi^\infty)\, e^{-\lambda(t - t_0)}.
\end{equation}

Having defined the growth rate, the GPU throughput at time $t$ is given by:
\begin{equation}
\label{eq:gpu_evolution}
\phi_{\text{GPU}}(t)=\phi_0\prod_{\tau}(1+r_\phi(\tau)),  
\end{equation}
where $\phi_0 = \phi_{\text{GPU}}(t_0)$, and $r_\phi(\tau)$ is the rate of compute growth at time $\tau$.
Eq.~\eqref{eq:gpu_evolution} formalizes this projection by accumulating compute gains from the B300 baseline year $t_0=2025$, with the same decelerating-rate form applied to memory bandwidth ($B_{mem} (t)$) through $r_B(t)$.

\begin{table}[width=0.99\textwidth, pos=bh]
\caption{This table summarizes the
GPU-generation inputs used for calibration, including peak and sustained compute, HBM bandwidth, HBM capacity,}
\label{tab:gpu_evolution}
\centering\footnotesize
\begin{tabular}{@{}lccccc@{}}
\toprule
& \textbf{A100} & \textbf{H100} & \textbf{B200} & \textbf{B300} & \textbf{CAGR} \\
& (2020) & (2022) & (2024) & (2025) & \\
\midrule
Peak FP16 (TFLOPS)  & 312  & 990   & 2{,}250 & 3{,}750 & 64\%/yr \\
\quad Sustained$^\dagger$ & 125  & 396   & 900  & 1{,}500 & 64\%/yr \\
Peak HBM (GB/s)     & 2{,}039 & 3{,}350 & 8{,}000 & 8{,}000 & 31\%/yr \\
\quad Sustained$^\dagger$ & 1{,}427 & 2{,}345 & 5{,}600 & 5{,}600 & 31\%/yr \\
HBM capacity (GB)   & 80   & 80    & 192     & 288    & --- \\
TDP (W)             & 400  & 700   & 1{,}000 & 1{,}400 & --- \\
\bottomrule
\multicolumn{6}{@{}l}{\footnotesize $^\dagger$Sustained = 40\% compute, 70\% bandwidth utilization (single-request inference).}
\end{tabular}
\end{table}

 Table~\ref{tab:gpu_evolution} calibrates these rates against NVIDIA A100--B300 2020--2025 CAGRs: $\sim$64\%/yr compute, $\sim$31\%/yr memory bandwidth.
The fitted deceleration model sets $r_\phi^0=64$\%/yr and $r_B^0=31$\%/yr, approaches $r_\phi^\infty=10$\%/yr and $r_B^\infty=7$\%/yr, and uses $\lambda=0.15$ for a slowdown half-life of $\sim$4.6~years.
 This calibration matches observed B200 $\sim$2.3$\times$ and B300 $\sim$3.8$\times$ H100 performance in 2024--2025, then scales the B300 (2025) baseline by $\sim$7$\times$ in 2030 and $\sim$56$\times$ in 2040.

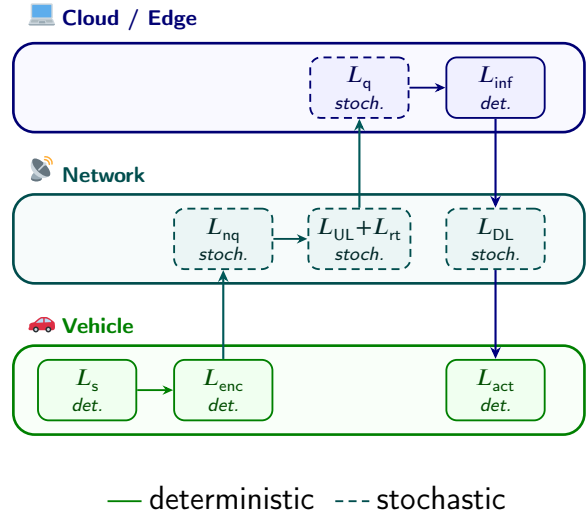
\begin{figure}[width=0.99\textwidth, pos = t]
    \centering
%

\colorlet{CloudFill}{blue!6}
\colorlet{CloudBorder}{blue!45!black}
\colorlet{CloudText}{blue!30!black}

\colorlet{NetFill}{teal!8}
\colorlet{NetBorder}{teal!60!black}
\colorlet{NetText}{teal!30!black}

\colorlet{VehFill}{green!6}
\colorlet{VehBorder}{green!50!black}
\colorlet{VehText}{green!30!black}

\colorlet{ArrowUp}{teal!70!black}
\colorlet{ArrowDown}{blue!50!black}
\colorlet{ArrowInt}{green!50!black}
\colorlet{LinkLabelCol}{gray!60!black}

\tikzset{
  detnode/.style={
    rectangle, rounded corners=4pt,
    minimum width=1.3cm, minimum height=0.8cm,
    align=center, font=\small,
    line width=0.7pt
  },
  stocnode/.style={
    detnode,
    dash pattern=on 3pt off 2pt
  },
  clouddet/.style={detnode,  fill=CloudFill, draw=CloudBorder, text=CloudText},
  cloudsoc/.style={stocnode, fill=CloudFill, draw=CloudBorder, text=CloudText},
  netdet/.style={detnode,  fill=NetFill, draw=NetBorder, text=NetText},
  netsoc/.style={stocnode, fill=NetFill, draw=NetBorder, text=NetText},
  vehdet/.style={detnode,  fill=VehFill, draw=VehBorder, text=VehText},
  vehsoc/.style={stocnode, fill=VehFill, draw=VehBorder, text=VehText},
  layerbg/.style={
    rectangle, rounded corners=8pt,
    inner sep=5pt, line width=1pt
  },
  layertag/.style={
    font=\footnotesize\bfseries,
    anchor=south west,
    inner sep=2pt
  },
  arrup/.style={
    -{Stealth[length=4.5pt,width=4pt]},
    line width=0.8pt, color=ArrowUp
  },
  arrdown/.style={
    -{Stealth[length=4.5pt,width=4pt]},
    line width=0.8pt, color=ArrowDown
  },
  arrint/.style={
    -{Stealth[length=4pt,width=3.5pt]},
    line width=0.7pt, color=ArrowInt
  },
  lnklbl/.style={
    font=\scriptsize, color=LinkLabelCol,
    inner sep=1.5pt, fill=white, fill opacity=0.85, text opacity=1
  },
  nullnode/.style={
    inner sep=0pt, outer sep=0pt,
    minimum size=0pt, draw=none, fill=none},
}
\newcommand{\layerheader}[4]{%
  \node[layertag, color=#3, xshift=4pt, yshift=2pt]
    at (#4.north west)
    {\resizebox{0.35cm}{!}{\twemoji{#1}}\;{\sffamily #2}};%
}

\begin{tikzpicture}[node distance=0pt]

\coordinate (xL) at (0,0);
\coordinate (xR) at (7.2,0);

\node[vehdet] (sense) at (0.8, 0)
  { $L_{\text{s}}$\\[-1pt]\scriptsize\textit{det.}};

\node[vehdet] (encode) at (2.6, 0)
  { $L_{\text{enc}}$\\[-1pt]\scriptsize\textit{det.}};

\node[vehdet] (actuate) at (6.2, 0)
  { $L_{\text{act}}$\\[-1pt]\scriptsize\textit{det.}};

\node[nullnode] (vehNull-L) at (xL) {};
\node[nullnode] (vehNull-R) at (xR) {};

\begin{scope}[on background layer]
  \node[layerbg, fill=VehFill!40, draw=VehBorder,
        fit=(vehNull-L)(sense)(encode)(actuate)(vehNull-R)] (vehLayer) {};
\end{scope}
\layerheader{automobile}{Vehicle}{VehBorder}{vehLayer}

\node[netsoc] (nq) at (2.6, 2.0)
  { $L_{\text{nq}}$\\[-1pt]\scriptsize\textit{stoch.}};

\node[netsoc] (ul) at (4.4, 2.0)
  { $L_{\text{UL}}{+}L_{\text{rt}}$\\[-1pt]\scriptsize\textit{stoch.}};

\node[netsoc] (dl) at (6.2, 2.0)
  { $L_{\text{DL}}$\\[-1pt]\scriptsize\textit{stoch.}};

\node[nullnode] (netNull-L) at (0, 2.0) {};
\node[nullnode] (netNull-R) at (7.2, 2.0) {};

\begin{scope}[on background layer]
  \node[layerbg,
        fill=NetFill!40, draw=NetBorder,
        fit=(netNull-L)(nq)(ul)(dl)(netNull-R)] (netLayer) {};
\end{scope}
\layerheader{satellite antenna}{Network}{NetBorder}{netLayer}

\node[cloudsoc] (queue) at (4.4, 4.0)
  { $L_{\text{q}}$\\[-1pt]\scriptsize\textit{stoch.}};

\node[clouddet] (infer) at (6.2, 4.0)
  { $L_{\text{inf}}$\\[-1pt]\scriptsize\textit{det.}};

\node[nullnode] (cloudNull-L) at (0, 4.0) {};
\node[nullnode] (cloudNull-R) at (7.2, 4.0) {};

\begin{scope}[on background layer]
  \node[layerbg,
        fill=CloudFill!40, draw=CloudBorder,
        fit=(cloudNull-L)(queue)(infer)(cloudNull-R)] (cloudLayer) {};
\end{scope}
\layerheader{laptop}{Cloud / Edge}{CloudBorder}{cloudLayer}

\draw[arrint] (sense.east) -- (encode.west);

\draw[arrint, color=CloudBorder]
  (queue.east) -- (infer.west);

\draw[arrup]
  (encode.north) -- (nq.south);

\draw[arrint, color=NetBorder]
  (nq.east) -- (ul.west);

\draw[arrup]
  (ul.north) -- (queue.south);

\draw[arrdown]
  (infer.south) -- (dl.north);

\draw[arrdown]
  (dl.south) -- (actuate.north);

\node[anchor=north west, font=\scriptsize, align=left,
      text=black!70!black,
      below=0.55cm of vehLayer.south, xshift=4pt] (leg) {%
  \tikz[baseline=-1ex]{\draw[VehBorder, line width=0.7pt]
    (0,0)--(0.45,0);}~{\large deterministic}\quad
  \tikz[baseline=-1ex]{\draw[NetBorder, line width=0.7pt,
    dash pattern=on 3pt off 2pt] (0,0)--(0.45,0);}~{\large stochastic}
};

\end{tikzpicture}
    \caption{End-to-end control loop delay decomposition. Solid borders denote deterministic components; dashed borders denote stochastic components that consume the remaining latency slack.}
    \label{fig:latency_pipeline}
\end{figure}

\subsection{Latency-Budget Decomposition}
\label{sec:stochastic_latency}

\begin{table}[width=\textwidth, pos=b]
\caption{Representative latency components in the control loop. Deterministic terms set the baseline floor, while stochastic terms consume the remaining slack under shared network and cloud loading.}
\label{tab:latency_components}
\centering
\footnotesize
\begin{tabular}{@{}>{\centering\arraybackslash}p{1.3cm}>{\centering\arraybackslash}p{1.1cm}p{1.4cm}p{11.0cm}@{}}
\toprule
\textbf{Component} & \textbf{Phase} & \textbf{Type} & \textbf{Interpretation} \\
\midrule
$L_{\text{s}}$ & Capture & Deterministic & Camera exposure, LiDAR scan accumulation, and sensor timestamp alignment. \\
$L_{\text{enc}}$ & Capture & Deterministic & Strategy-dependent local preprocessing: S1 video compression, S2 backbone features, S3 encoder/query generation. \\
$L_{\text{nq}}$ & Offload & Stochastic  & Uplink scheduling delay that grows rapidly with cell loading and interference. \\
$L_{\text{UL}}$ & Offload & Stochastic & Air-interface transmission and HARQ retransmission; generation dependent. \\
$L_{\text{rt}}$ & Offload & Stochastic & Fiber propagation, switching, and backhaul transport from gNB to the assigned edge site. \\
$L_{\text{q}}$ & Compute & Stochastic  & Waiting time for an available GPU, bounded via the provisioning model in Section~\ref{sec:gpu_service_model}. \\
$L_{\text{inf}}$ & Compute & Deterministic  & Approximately 1~ms (E2E), 58~ms (VLM), and 145~ms (VLA) for feature-level offloading on the B300 (2025) baseline, using the Section~\ref{sec:gpu_service_model} roofline calibration and declining with GPU evolution. \\
$L_{\text{DL}}$ & Return & Stochastic & Downlink delivery of control or trajectory outputs; small payload but still scheduling limited. \\
$L_{\text{act}}$ & Return & Deterministic  & Safety checks, trajectory post-processing, and CAN bus actuation. \\
\bottomrule
\end{tabular}
\end{table}

The decomposition uses Section~\ref{sec:comm_model} and Section~\ref{sec:gpu_service_model} to separate deterministic and stochastic latency terms. Formally, the end-to-end latency is the sum of the nine deterministic and stochastic phase terms:
\begin{equation}
\label{eq:latency}
L_{\text{total}} = \underbrace{L_{\text{s}} + L_{\text{enc}}}_{\text{capture}} + \underbrace{L_{\text{nq}} + L_{\text{UL}} + L_{\text{rt}}}_{\text{offload}} + \underbrace{L_{\text{q}} + L_{\text{inf}}}_{\text{compute}} + \underbrace{L_{\text{DL}} + L_{\text{act}}}_{\text{return}}.
\end{equation}

The equation groups the nine terms into capture, offload, compute, and return phases. In this organization, Figure~\ref{fig:latency_pipeline} marks deterministic per-cycle terms with solid borders and stochastic communication and GPU contention terms with dashed borders. Table~\ref{tab:latency_components} summarizes all latency components, their values, and their interpretations.

\subsubsection{Deterministic Latency Floor}
For latency decomposition, \textbf{sensing delay ($L_{\text{s}}$)} isolates the fixed local latency that precedes vehicle-to-network communication and cloud inference. We model this fixed sensing delay as a 5~ms term. The term includes camera exposure, LiDAR scan accumulation, and timestamp alignment. Together, these operations form a fixed prerequisite that is independent of cloud offloading. Quantitatively, this fixed sensing delay consumes 5\% of the reactive 100~ms control-loop budget (and only 1.7\% of the 300~ms deliberative budget defined below). Feasibility depends on whether vehicle-to-network communication and cloud inference fit the remaining slack.

We quantify \textbf{encoding delay ($L_{\text{enc}}$)}, the onboard computation incurred by each offloading strategy before vehicle-to-network communication and cloud inference. $L_{\text{enc}}$ measures the local latency each strategy adds to reduce uplink demand. For the lighter strategies, $S1$ adds H.265 compression delay, while $S2$ runs the vision backbone onboard, raising local delay to 5--8~ms but reducing uplink demand. Under $S3$, the transformer encoder also runs onboard, increasing the delay to 8--12~ms.

\textbf{Cloud inference delay $L_{\text{inf}}$} Once the request reaches an available GPU, the GPU Service Model in Section~\ref{sec:gpu_service_model} assigns the offloaded workload a cloud inference delay. Accordingly, $L_{\text{inf}}$ is the time that it takes the assigned GPU to process the input and predict the right action.

\textbf{Actuation delay $L_{\text{act}}$} is post-return delay. We denote this local actuation delay by $L_{\text{act}}$. This term is fixed at 3~ms after the cloud result returns. This 3 ms term covers trajectory post-processing, safety checks, and Controller Area Network (CAN) bus command issuance. Thus, $L_{\mathrm{act}}$ captures the residual onboard safety and execution layer. Because $L_{\mathrm{act}}$ is hardware-determined, we keep it constant across scenarios and include it in the deterministic delay floor.

\subsubsection{Stochastic Latency Components}

\textbf{Network scheduling delay ($L_{\text{nq}}$)} is stochastic and can range from negligible values to more than 50~ms under heavy contention. When $n$ vehicles share a cell, each receives $1/n$ of uplink resources, increasing frame transmission time. We model this via a processor-sharing (PS) queue approximation~\citep{borst2007dimensioning}. This is the component that turns communication scarcity into a nonlinear systems bottleneck: as loading rises, vehicles not only split bandwidth but also degrade each other's spectral efficiency through interference. The frame size is $D_s = B_s / f$ bits and the cell loading ratio is
\begin{equation}
\label{eq:loading_ratio}
\rho_c(n, s, g) = \frac{n \cdot B_s}{C_c(n,g)},
\end{equation}
where $C_c(n,g) = W_c^{(g)} \cdot \eta_v(n,g)$ is the interference-degraded cell capacity.
Both numerator and denominator depend on $n$; the denominator decreases with $n$ due to interference (Eq.~\eqref{eq:spec_eff}), so $\rho_c$ grows faster than linearly. The mean scheduling delay (excess above unloaded service time) is
\begin{equation}
\label{eq:scheduling_delay}
\mathbb{E}[L_{\text{nq}}] = \frac{D_s}{C_c(n,g)} \cdot \frac{\rho_c(n,s,g)}{1 - \rho_c(n,s,g)}.
\end{equation}

This is zero at $n = 1$ and grows sharply as $\rho_c \to 1$, creating the capacity cliff that later dominates the dense urban case study. The PS sojourn time has an exponential tail~\citep{borst2007dimensioning}:
\begin{equation}
\label{eq:scheduling_tail}
\Pr(L_{\text{nq}} > l) \leq \frac{\rho_c(n,s,g)}{1 - \rho_c(n,s,g)} \exp\!\left(-\mu_c (1 - \rho_c(n,s,g))\, l\right),
\end{equation}
where $\mu_c = C_c / D_s$ is the cell service rate.

\textbf{Uplink transmission delay ($L_{\text{UL}}$)} is stochastic and typically falls in the 0.1--4~ms range. It represents the air-interface time from vehicle to gNB, including channel coding and hybrid automatic repeat request (HARQ) retransmission. Although smaller than the cloud inference time, $L_{\text{UL}}$ links the communication generation to slot duration and retransmission behavior, which motivates modeling it explicitly. We capture this stochastic air-interface component with a shifted-exponential model: \begin{equation}\label{eq:ul_delay}L_{\text{UL}} \sim L_{\text{UL}}^{\min} + \mathrm{Exp}(\lambda_{\text{UL}}^{(g)}).\end{equation}
The equation parameterizes this planning-cycle upload delay as one generation-dependent slot duration, $L_{\text{UL}}^{\min}$ (1.0~ms for 5G, 0.5~ms for 5G-Advanced, and 0.1~ms for 6G), plus an exponential random component whose rate $\lambda_{\text{UL}}^{(g)}$ captures HARQ and scheduling variability. 

\textbf{Routing and backhaul delay ($L_{\text{rt}}$)} is the delay between the gNB and assigned edge site. This delay captures random fiber-backhaul transport from the gNB to the assigned edge site, usually about 0.5--10~ms.
 Here, backhaul denotes transport between the gNB and assigned edge site. For assigned edge sites within 150~km, fiber propagation is only about 1~ms, so switching and transport queueing drive placement-sensitive delay. Backhaul delay makes edge-site placement part of the closed-loop latency constraint. 
 We therefore treat $L_{\text{rt}}$ as a placement-sensitive random delay before adding GPU queueing delay at the assigned edge site. Thus, $L_{\text{rt}}$ is shifted-exponential with mean 5~ms for 5G, 2~ms for 5G-Advanced with optimized MEC placement, and 0.5~ms for 6G with integrated access-backhaul. 

\textbf{GPU queueing delay ($L_{\text{q}}$)} ranges from near zero to about 5~ms before GPU service begins. We estimate this wait with an M/D/$c_e$ queue and the Erlang-C tail bound from Section~\ref{sec:gpu_service_model}. This tail bound converts GPU sharing into a site-level constraint on rare long-wait events. Solving $\Pr(L_{\text{q}}>L_{\text{q}}^{\max}) \leq \epsilon_q$ gives the over-provisioning above mean load needed to meet each site-level queueing budget. At $\epsilon_q=10^{-5}$, roughly 15--25\% over-provisioning fixes queueing at typical loads (less at the densest sites) before cloud inference.

\textbf{Downlink transmission delay ($L_{\text{DL}}$)} happens after cloud inference. This downlink carries trajectory or control commands in about 1--2~ms. Because payloads are $<$1~Mbps, downlink delay is scheduling-dominated. This scheduling-dominated behavior, rather than payload size, motivates using the same shifted-exponential form for $L_{\text{DL}}$. We model $L_{\text{DL}}$ with a shifted exponential form analogous to $L_{\text{UL}}$.

\begin{table}[t]
\caption{Stochastic communication-delay parameters used in the latency-tail calculation. Minimum delays are deterministic floors inside shifted-exponential components; means are total component means.}
\label{tab:stochastic_comm_latency_params}
\centering
\footnotesize
\begin{tabular}{@{}lccc@{}}
\toprule
\textbf{Parameter} & \textbf{5G} & \textbf{5G-Adv} & \textbf{6G} \\
\midrule
$\lambda_{\text{UL}}^{(g)}$ (s$^{-1}$) & 500 & 2{,}000 & 10{,}000 \\
$L_{\text{UL}}^{\min}$ (ms) & 1.0 & 0.5 & 0.1 \\
$L_{\text{rt}}^{\min}$ (ms) & 1.0 & 0.5 & 0.1 \\
$\mathbb{E}[L_{\text{rt}}]$ (ms) & 5.0 & 2.0 & 0.5 \\
$L_{\text{DL}}^{\min}$ (ms) & 0.5 & 0.25 & 0.1 \\
$\mathbb{E}[L_{\text{DL}}]$ (ms) & 2.0 & 1.5 & 1.0 \\
\bottomrule
\end{tabular}
\end{table}

\subsubsection{Control-Loop Delay Budget}
Cloud driving must close a sense--plan--act loop, and the admissible loop time depends on whether the cloud result drives \emph{reactive} control or \emph{deliberative} planning. We therefore evaluate two budgets. The \textbf{reactive budget}, $L_{\max}=100$~ms, follows from the 10~Hz planning period and is the deadline a cloud decision must meet to steer the vehicle within the current cycle; we treat it as the primary budget. The \textbf{deliberative budget}, $L_{\max}=300$~ms, models a slower planning tier whose output refines, but does not gate, immediate control; it is admissible only when an onboard reactive controller guarantees a fallback within the 100~ms loop, consistent with the fast--slow safety hierarchies used in deployed AV stacks. Reporting both budgets isolates when latency---rather than bandwidth or cost---is the binding constraint. The reactive threshold is itself the binding parameter for the compute regime: at the 2025 baseline the VLA deterministic floor is 132--164~ms, so relaxing the budget to $\sim$150~ms would admit VLA already in 2025--2026, whereas at 100~ms the floor first clears in 2027. We therefore treat the near-term VLA compute-bound result as specific to the tight reactive deadline, and report the deliberative tier alongside it rather than choosing one in isolation.

Combining sensing, encoding, cloud inference, and the actuation delay gives the deterministic delay floor:
\begin{equation}
\label{eq:det_floor}
L_{\text{det}}(s, m, t) = L_{\text{s}} + L_{\text{enc}}(s) + L_{\text{inf}}(s, m, t) + L_{\text{act}}.
\end{equation}
Each budget turns that floor into a remaining stochastic budget for network and queueing delay:
\begin{equation}
\label{eq:stoch_budget}
\Delta(s, m, t)=L_{\max}-L_{\text{det}}(s, m, t).
\end{equation}

Put differently, deterministic per-cycle delays are paid first, and only the remaining budget can be used for stochastic scheduling, transmission, routing, and queueing delay. A positive remaining budget is therefore a prerequisite for any tail-latency check. If the remaining budget is nonpositive, deterministic delay already exhausts the loop budget, making the tail constraint infeasible regardless of communication capacity.

When the remaining budget is positive, the five stochastic components determine whether the scenario meets the tail target. Each stochastic component is modelled with a shifted exponential random variable with parameters in table~\ref{tab:stochastic_comm_latency_params}. Tail-latency feasibility constrains exceedance probability rather than mean delay, so we require the stochastic sum to fit inside the remaining budget with reliability $\epsilon=10^{-5}$:
\begin{equation}
\label{eq:stoch_tail}
p_{\text{tail}}(\bar{n}_{\text{cell}},s,g,m,t)
=
\Pr\!\left(
L_{\text{nq}}+L_{\text{UL}}+L_{\text{rt}}+L_{\text{q}}+L_{\text{DL}}
>
\Delta(s,m,t)
\right)
\leq \epsilon.
\end{equation}

\begin{figure}[width=0.99\textwidth, pos = t]
\centering
\includegraphics[width=0.55\linewidth]{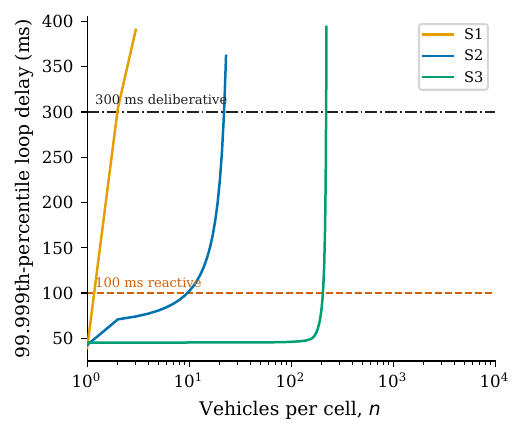}
\caption{Tail latency under increasing cell utilization. Curves show the 99.999th-percentile end-to-end loop delay under 5G-Advanced (compact E2E workload) as vehicles per cell increase for S1--S3. Horizontal lines mark the reactive 100~ms and deliberative 300~ms budgets. The sharp rise at high loading is the capacity cliff discussed in the text: stochastic network scheduling consumes the remaining budget~$\Delta$, so points above a budget violate Eq.~\eqref{eq:stoch_tail} and fail the latency side of the feasibility screen at that budget.}
\label{fig:latency_budget}
\end{figure}

The tail-probability expression combines network scheduling, uplink transmission, routing and backhaul, GPU queueing, and downlink transmission as independent stochastic delays within the remaining budget. We evaluate this exceedance probability with a Chernoff bound, using the moment-generating function (MGF) of each independent delay at each communication generation and cell loading. This calculation yields the maximum feasible cell loading for each communication generation and identifies which stochastic delay binds. At dense loading the binding term is network scheduling; in GPU-heavy scenarios it shifts to GPU queueing and inference. Figure~\ref{fig:latency_budget} makes this concrete: the 99.999th-percentile loop delay stays flat as vehicles per cell grow, then rises sharply through the reactive 100~ms and deliberative 300~ms budgets as network-scheduling delay exhausts the remaining budget~$\Delta$ and points above a budget violate Eq.~\eqref{eq:stoch_tail}.


\subsection{Joint Feasibility Gate}
\label{sec:feasibility}

Together, the communication model and latency model turn each scenario into a pair of technical gates. The communication gate checks whether the access cell can admit the expected active vehicles for the chosen offloading strategy and generation. The latency gate checks whether deterministic slack and stochastic tail compliance both hold under that same scenario. Only scenarios that pass both gates proceed to edge placement and cost comparison.

\subsubsection{Communication Gate}
Communication feasibility holds when the expected number of active vehicles per cell stays within the maximum cell capacity computed using the interference-aware rate model:

\begin{equation}
\label{eq:comm_feasibility}
\bar{n}_{\text{cell}} \leq N_c^{\max}(s,g).
\end{equation}

The admission rule is direct: aggregate uplink demand must fit within the cell's maximum uplink capacity before costs are considered. A branch that fails this rule is removed because the access network cannot carry its sustained uplink traffic. Thus, the communication gate acts as a capacity filter rather than an economic filter. Because aggregate demand is the number of active vehicles multiplied by the per-vehicle bitrate, high-bitrate strategies fail the filter first. The gate is therefore most restrictive for S1 under 5G, where a per-vehicle uplink requirement of 100~Mbps causes aggregate demand to reach the maximum cell capacity with few admitted vehicles.

\subsubsection{Latency Gate}
We evaluate latency feasibility with two checks: deterministic latency-budget sufficiency and stochastic tail compliance. The first check asks whether fixed sensing, encoding, inference, and actuation delays leave any budget for random delay. The second check evaluates the tail-probability metric for the scenario's cell loading, model class, GPU generation, and communication generation. The latency gate passes only when deterministic slack is positive and the tail probability is no larger than the target threshold.
If deterministic slack is nonpositive, the chosen budget leaves no room for random network or queueing delay; under the reactive 100~ms budget this is precisely what excludes near-term VLA inference regardless of communication capacity (Section~\ref{sec:cs_compute}). If deterministic slack is positive, high cell load can still violate the network-scheduling tail bound. Hence, feasibility requires both positive deterministic slack and tail compliance in the same scenario, evaluated separately under the reactive and deliberative budgets.

\subsubsection{Branch Feasibility Indicator}
Joint feasibility is summarized by a binary indicator that records the combined outcome of the communication and latency gates:
\begin{equation}
\label{eq:feasibility}
\psi(s,g,m,t)=
\begin{cases}
1, & \bar{n}_{\text{cell}}\leq N_c^{\max}(s,g),\ \Delta(s,m,t)>0,\ p_{\text{tail}}(\bar{n}_{\text{cell}},s,g,m,t)\leq\epsilon,\\
0, & \text{otherwise.}
\end{cases}
\end{equation}

An active indicator means that access capacity, positive deterministic slack, and bounded tail probability all hold simultaneously. When the indicator is active, the scenario remains eligible for optimization. When the indicator is inactive, the scenario is excluded from EIO assignment and provisioning decisions.

\subsection{Edge Infrastructure Optimization}
\label{sec:edge_placement}

Once a branch passes the communication and latency gates, the remaining technical question is where to place edge capacity. At that point, the queueing model has already determined the minimum GPU count needed to satisfy the queue-tail bound for assigned demand, so the infrastructure optimization chooses which sites to open and which vehicles to assign to those sites.
Feasibility sets the scope of this placement problem: inactive branches receive no vehicle assignments and no cloud resources, while active branches are evaluated through the capacitated facility location problem.

The facility-location decision set has three linked parts: site opening, vehicle assignment, and GPU count. Vehicle assignment determines each site's request arrival rate, and that arrival rate determines the minimum queue-safe GPU count. The objective and constraints encode this same chain from branch feasibility to assignment to provisioning:

\begin{equation}
\label{eq:facility_location}
\min_{x, y, c} \;\; \sum_{e \in \mathcal{E}} \left( F_e \cdot x_e + \sum_{v \in \mathcal{V}} a_{ve} \cdot y_{ve} + \gamma(t) \cdot c_e \right)
\end{equation}

\begin{align}
\sum_{e \in \mathcal{E}} y_{ve} &= \psi & \forall\, v \in \mathcal{V} & \quad \text{(feasible-demand assignment)} \label{eq:assign} \\
y_{ve} &\leq x_e & \forall\, v \in \mathcal{V},\; e \in \mathcal{E} & \quad \text{(open-site coupling)} \label{eq:open} \\
d_{ve} \cdot y_{ve} &\leq D_{\max}^{(g)} & \forall\, v \in \mathcal{V},\; e \in \mathcal{E} & \quad \text{(latency coverage)} \label{eq:coverage} \\
\sum_{v \in \mathcal{V}} y_{ve} \cdot f &\leq c_e \mu_{\text{eff}}(s,m,t) \rho_{\max} & \forall\, e \in \mathcal{E} & \quad \text{(GPU capacity)} \label{eq:gpu_cap} \\
c_e &\geq \psi \, c_e^*(s,m,t) & \forall\, e \in \mathcal{E} & \quad \text{(queue-tail provisioning)} \label{eq:tail_site}
\end{align}

The constraints follow the same order as the deployment logic. First, the activation and open-site coupling constraints require each vehicle to choose an opened site only when the branch is feasible. Next, the coverage constraint limits vehicle--site pairs to latency-reachable locations within the round-trip budget. Finally, the capacity and queue-tail constraints ensure that assigned demand fits provisioned GPU throughput and meets the rare-wait guarantee.
After those feasibility, coverage, and provisioning constraints are enforced, the objective adds the costs of the chosen deployment: facility fixed cost, assignment and backhaul cost, and annualized GPU cost. This produces the mixed-integer nonlinear program that we solve by greedy rounding for the New York City deployment.

\subsection{Cost Crossover}
\label{sec:tco}

The final economic comparison follows feasibility screening, queueing and inference latency estimation, and edge assignment. At this point, the remaining question is economic rather than technical: whether a feasible cloud branch costs less per vehicle than the hardware it avoids. Accordingly, the cost comparison includes only offloading options that satisfy communication and latency feasibility.

For each feasible offloading option, per-vehicle cloud cost aggregates EIO deployment costs and allocates them over active demand at utilization. The cloud-cost numerator combines annualized GPU provisioning, fixed site cost, assignment and backhaul cost, and strategy-dependent network cost; the utilization-adjusted denominator spreads that numerator across active vehicles. The onboard-hardware terms then annualize, over the vehicle lifetime, the full-pipeline and residual onboard system-on-chip compute, both priced at the model-class unit cost $\beta_m(t)$. Together, the cloud-cost side and the avoided-hardware side define the feasible-option TCO comparison:
\begin{equation}
\label{eq:crossover}
C_{\text{cloud}}^{\text{per-veh}}(s,m,u,t)
\leq
\beta_m(t)\left(H_m^{\text{full}}-H_s\right).
\end{equation}

The crossover inequality applies that comparison directly: feasible offloading is preferred when per-vehicle cloud cost does not exceed the avoided onboard hardware cost $\beta_m(t)\left(H_m^{\text{full}}-H_s\right)$, i.e.\ the model-class cost of the compute moved to the cloud. Because cloud cost varies with fleet utilization while avoided onboard hardware is fixed per vehicle, utilization governs the cost crossover. Once communication and latency are feasible, this makes fleet utilization the primary cost-crossover driver. Thus, cloud offloading favors low- and moderate-utilization fleets sharing GPUs rather than replicating onboard hardware.

Table~\ref{tab:case_cost_inputs} details our cost assumptions. More specifically, it states the projected price declines for both in-vehicle and cloud-side GPUs and introduces the infrastructure cost symbols used in equation~\ref{eq:facility_location}. The key accounting asymmetry is that onboard hardware is purchased for every vehicle, while cloud GPU capacity is provisioned for active demand, so utilization becomes central after communication feasibility is satisfied.

\begin{table}[h!]
\caption{Reference cost inputs for the case-study comparisons.}
\label{tab:case_cost_inputs}
\centering
\footnotesize
\begin{tabular}{@{}p{3.2cm}p{3.2cm}p{5.2cm}@{}}
\toprule
\textbf{Item} & \textbf{2026 reference} & \textbf{Interpretation} \\
\midrule
\multicolumn{3}{@{}l}{\textbf{In-vehicle baselines}} \\
Compact E2E baseline & \$400/vehicle-year & Orin-class onboard hardware, declining 8\%/yr. \\
VLM baseline & \$1,000/vehicle-year & Thor-class onboard hardware from 2026, declining 10\%/yr. \\
VLA baseline & \$8,500/vehicle-year & RTX PRO 6000-class onboard hardware~\citep{nvidia2025rtxpro6000}, declining 15\%/yr. \\
\midrule
\multicolumn{3}{@{}l}{\textbf{Cloud cost components}} \\
GPU provisioning & $\gamma(t) c_e$, initially \$10K/yr per GPU & Annualized hardware cost with $c_e$ constrained by Eq.~\eqref{eq:min_gpus} and declining with Eq.~\eqref{eq:gpu_evolution}. \\
Facility cost & $F_e(t)$ & Fixed annualized site cost for power, space, and interconnection. \\
Network cost & $\kappa_s(t)$ & Strategy-dependent backhaul and transport cost. \\
Effective cloud cost decline & $\sim$44\%/yr near-term; $\sim$25\%/yr by 2035 & Combines GPU price reduction with performance improvement from Eq.~\eqref{eq:gpu_evolution}. \\
\bottomrule
\end{tabular}
\end{table}

\section{Case Studies}
\label{sec:case_studies}

The case study applies the analytical pipeline to a single dense metropolitan setting---New York City---and reads it through three nested binding regimes. Rather than contrasting two geographies, we hold one scenario fixed and ask which constraint binds: the access uplink (Section~\ref{sec:cs_comm}), the reactive latency budget (Section~\ref{sec:cs_compute}), or per-vehicle cost (Section~\ref{sec:cs_cost}). The regimes are sequential. A branch must clear the uplink before latency matters, and must be latency-admissible before its cost is worth comparing. Cloud driving is therefore not simply feasible or infeasible in general; the binding factor changes with model class, GPU year, and which latency budget applies.

We use the scenario-input grid from Section~\ref{sec:scenario_inputs}. NYC has $\sim$2.2M registered vehicles across $\sim$10{,}000 cell sites; per-cell uplink bandwidth is 50~MHz for 5G FR2, 100~MHz for 5G-Advanced, and 400~MHz for 6G, with interference-aware spectral efficiencies of 5, 7, and 20~bps/Hz at single-user loading~\citep{3gpp2021ts38300, garcia2021tutorial, andrews2011tractable, 3gpp2020tr38901}; the 6G values follow ITU IMT-2030 targets~\citep{itu2023imt2030}. The scenario matrix is a Cartesian product of three communication generations, three model classes, three offloading strategies, eight AV penetration levels $\{0.1\%, 1\%, 5\%, 10\%, 20\%, 30\%, 50\%, 100\%\}$, and six utilization levels $u \in \{0.05, 0.12, 0.30, 0.45, 0.65, 1.0\}$; thus, the study has $3 \times 3 \times 3 \times 8 \times 6 = 1{,}296$ cloud-offloading branches per reference year (full in-vehicle inference is the cost baseline, not a cloud branch). The communication gate alone leaves 984 uplink-feasible and 312 uplink-infeasible branches. Adding the reactive 100~ms latency gate reduces the jointly feasible set to 487 branches in 2026; under the 300~ms deliberative budget, 903 branches are jointly feasible. The gap between these two counts is the compute-bound regime made quantitative.

The three offloading strategies organize the evidence. S1, raw-sensor offloading, moves the most computation to the cloud but creates the largest uplink burden. S3, query-level offloading, is easiest for the network but leaves substantial onboard hardware in every vehicle. S2, feature-level offloading, keeps the vision backbone local and sends the expensive prediction and planning stages to the cloud.

\subsection{Regime 1: Communication-Bound --- Dense Urban Cells}
\label{sec:cs_comm}

The NYC case isolates uplink feasibility under dense-cell loading. NYC has $\sim$2.2M registered vehicles across $\sim$10,000 cell sites,\footnote{NY DMV Statistical Data, \url{https://dmv.ny.gov/records/statistical-data}. Cell site count estimated from FCC ASR database.} and AV penetration is measured as a share of total registered vehicles. Thus 1\%, 5\%, 10\%, and 20\% penetration correspond to about 22K, 110K, 220K, and 440K AVs. At the reference utilization $u=0.45$, the 10\% point yields $\bar{n}_{\text{cell}}\approx10$ active vehicles per cell on average. That loading makes the per-vehicle uplink share, the bandwidth each vehicle receives after the cell is divided among active users, the first binding constraint rather than cloud GPU cost. The NYC analysis therefore asks a narrow pre-TCO question: which model, strategy, and communication-generation combinations can even reach the cloud?

\begin{figure}[width=\textwidth, pos = !t]
\centering
\includegraphics[width=\linewidth]{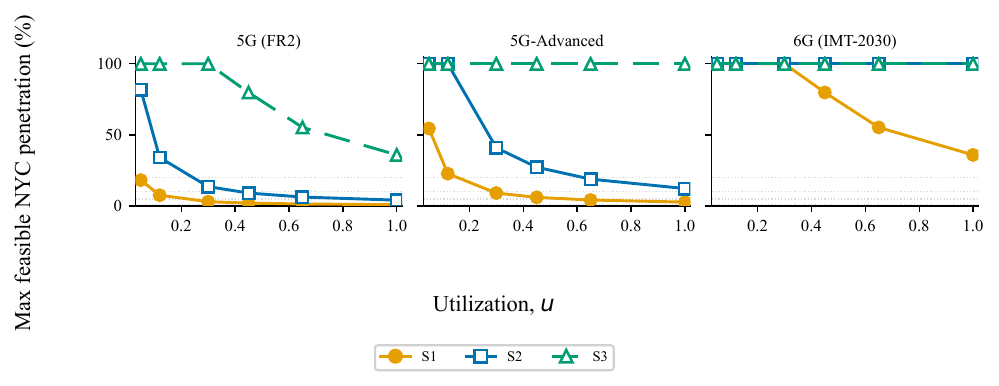}
\caption{Case Study 1: NYC communication frontier. The y-axis gives the maximum citywide fleet penetration that remains bandwidth-feasible at each utilization. Reference points: 1\% = 22K vehicles, 10\% = 220K vehicles.}
\label{fig:urban_frontier}
\end{figure}

\begin{figure}[width=0.99\textwidth, pos = t]
\centering
\includegraphics[width=0.98\linewidth]{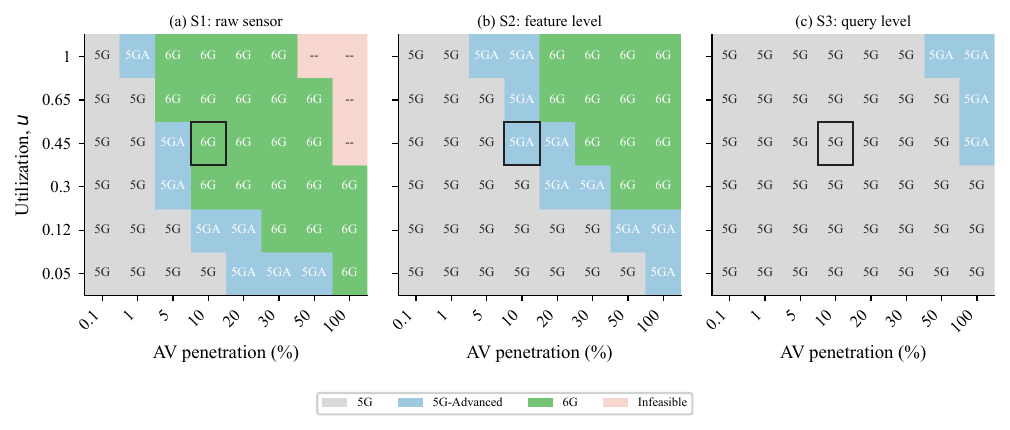}
\caption{Case Study 1: NYC communication sweep. Each cell shows the minimum communication generation required to keep a strategy bandwidth-feasible for the given AV penetration and utilization pair. The highlighted cell corresponds to the reference point of 10\% penetration and $u = 0.45$.}
\label{fig:nyc_generation_requirement}
\end{figure}

Figure~\ref{fig:urban_frontier} shows how dense loading becomes an uplink bottleneck. Each strategy has a different per-vehicle uplink requirement: 100~Mbps for S1, 25~Mbps for S2, and 3~Mbps for S3. As more AVs share the same cell, bandwidth is divided across users and interference-aware spectral efficiency declines, as captured by Eq.~\eqref{eq:per_user_bandwidth}. Once active loading exceeds $N_c^{\max}$, the branch is uplink-infeasible no matter which model class is used or how cheap cloud GPUs become. At $u=0.45$, 5G supports only about 2\% penetration for S1 and about 9\% for S2, so a 10\% citywide AV fleet already exceeds the 5G uplink limit for feature-level offloading. 5G-Advanced shifts the S2 feasibility boundary into the 10\%--30\% range, and 6G adds the headroom required for broader S2 and S3 deployment.

The operational lesson is that dense urban feasibility is controlled by concurrent vehicles per cell, not just total citywide fleet size. A deployment can look feasible on a citywide average and still fail in a Manhattan corridor, at a major intersection, or during a rush-hour hotspot. That is why the NYC result is a communication screen before it is an economic screen. Branches such as VLA + S1 + 5G are not plausible dense-cell cloud-driving candidates because they combine the most demanding model class, the most uplink-heavy strategy, and the weakest communication generation. Even if cloud inference were cheap, that branch cannot be supported reliably enough to matter.

Figure~\ref{fig:nyc_generation_requirement} presents the same result as a minimum-generation requirement map and makes S2's role easier to see. Under S1, most of the penetration-utilization grid requires 6G, and the most aggressive cells remain infeasible even with the assumed 6G capacity. Under S3, most cells are easy for the network, but that communication advantage is bought by keeping much more computation on the vehicle. S2 is the useful middle: most penetration and utilization combinations fit within the 6G envelope, and many moderate combinations are already reachable with 5G-Advanced. Because the figure screens communication feasibility only, it should be read as the access-network prerequisite for cloud driving, not as a complete cost result. \textbf{At urban cell density, communication generation is the binding variable, and S2 is the most useful planning target because it reduces uplink demand without giving up the main cloud-compute savings.}

\subsection{Regime 2: Compute-Bound --- Near-Term VLA under the Reactive Budget}
\label{sec:cs_compute}

The communication regime asks whether the uplink can carry a branch. The compute regime asks a question the deliberative budget hides: even with adequate bandwidth and an idle GPU, can the cloud return a VLA decision within the reactive 100~ms loop? For E2E and VLM workloads, the answer is yes throughout the study period, because their inference floor is small. For VLA it is not, and the reason is structural rather than economic.

\begin{figure}[width=\textwidth, pos = t]
\centering
\begin{subfigure}{0.49\linewidth}
\centering
\includegraphics[width=\linewidth]{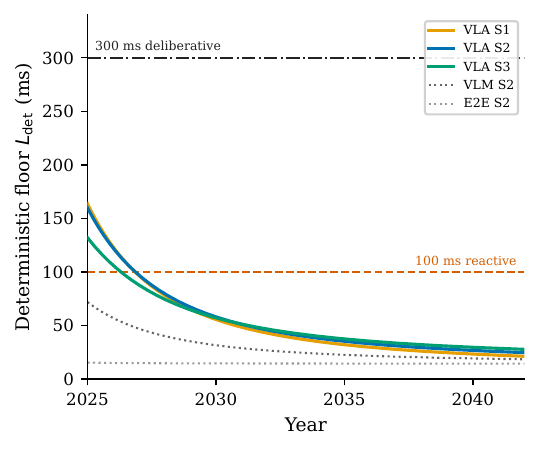}
\caption{Latency admissibility by year}
\end{subfigure}\hfill
\begin{subfigure}{0.49\linewidth}
\centering
\includegraphics[width=\linewidth]{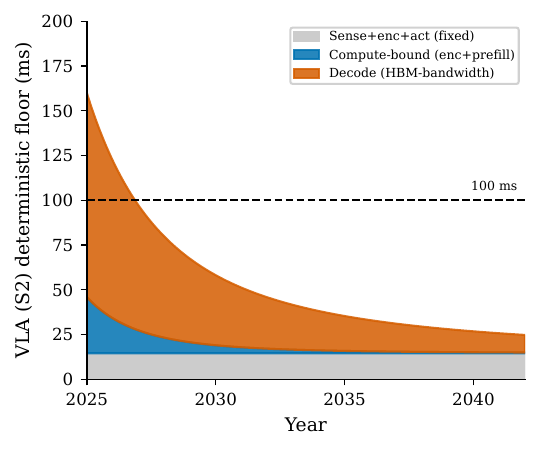}
\caption{Why VLA is memory-bound}
\end{subfigure}
\caption{Regime 2 (compute-bound). (a) Deterministic latency floor $L_{\text{det}}$ versus GPU year for VLA (S1--S3), with VLM-S2 and E2E-S2 references; the reactive 100~ms and deliberative 300~ms budgets are marked. The VLA \emph{floor} first drops below 100~ms in 2027, but this is a lower bound: once the access-scheduling tail is added, reactive admissibility is later and generation-dependent (Section~\ref{sec:cs_compute}); the floor is under 300~ms from 2025. (b) Decomposition of the VLA-S2 floor: the decode term (autoregressive reasoning and trajectory tokens) is HBM-bandwidth-bound and dominates the wall, so it shrinks only with memory bandwidth, not arithmetic throughput.}
\label{fig:compute_bound_floor}
\end{figure}

Figure~\ref{fig:compute_bound_floor}(a) plots the deterministic latency floor $L_{\text{det}}$---sensing, encoding, cloud inference, and actuation, with no network or queueing delay---against GPU year. On 2025-class hardware (B300) the VLA floor is 132--164~ms across S1--S3, above the 100~ms reactive budget for every offloading strategy and communication generation. The inference term dominates: Figure~\ref{fig:compute_bound_floor}(b) decomposes the VLA-S2 floor into a compute-bound part (encoder and prefill, which shrinks with GPU throughput) and a decode part (autoregressive reasoning and trajectory tokens, which shrinks only with HBM bandwidth). Decode alone is $\sim$114~ms in 2025 and falls slowly, because each generated token re-reads the model weights from memory; the workload is memory-bandwidth-bound, not arithmetic-bound. Within the FP16, dense, single-request autoregressive stack we calibrate (Alpamayo on TensorRT-LLM), the binding term is therefore HBM bandwidth: neither extra uplink bandwidth nor cheaper GPUs makes near-term VLA reactive-feasible---only growth in memory bandwidth does. This wall is stack-specific. FP8 weights roughly halve the per-token read ($\sim$114~ms~$\to$~$\sim$57~ms on 2025 hardware), and a parallel diffusion or flow action decoder removes the autoregressive re-reads entirely (decode falling to tens of milliseconds); either would move VLA's reactive-budget admissibility several years earlier. We report the conservative autoregressive FP16 case as a latency upper bound.

The deterministic floor first clears 100~ms in 2027, but the floor is only a lower bound; the stochastic tail must also fit. Counting jointly feasible branches per model class (of $3\times3\times8\times6=432$ each) makes the gate concrete. Under the reactive budget, VLA is feasible in $0/432$ branches in 2026, $119/432$ in 2028, and $204/432$ in 2030, while E2E sits near $283/432$ throughout and VLM sits at $204/432$ in 2026, rising to $237/432$ by 2030; these three 2026 counts ($283+204+0$) recover the $487$ jointly feasible branches reported in Section~\ref{sec:case_studies}. Under the deliberative budget the VLA count is $293/432$ already in 2026 (with E2E at $307/432$ and VLM at $303/432$, totaling the $903$ deliberative branches): the 300~ms tier effectively switches the VLA latency gate off. The communication generation then decides whether and when a loaded cell also clears the scheduling tail. At the reference dense corridor (10\% penetration, $u=0.45$, $\sim$10 vehicles per cell), 6G admits VLA-S2 by $\sim$2028, whereas 5G-Advanced's far heavier scheduling tail leaves that corridor reactive-infeasible across the studied horizon; only at light cell loading does 5G-Advanced admit VLA-S2, and even then not before $\sim$2029. \textbf{Under the reactive budget, latency---the memory-bandwidth-bound VLA decode together with the access-scheduling tail---decides which model class is admissible in which year and under which generation, largely independent of GPU cost. The deliberative tier removes this gate, but only for vehicles that carry an onboard reactive fallback.}

\subsection{Regime 3: Cost-Bound --- Utilization Economics}
\label{sec:cs_cost}

Once a branch is communication- and latency-admissible, the remaining question is economic: does utilization-pooled cloud inference cost less per vehicle than the onboard hardware it replaces? Here NYC's $\sim$2.2M registered vehicles are an asset rather than an obstacle. Cloud GPUs are provisioned for concurrent active demand ($Nu$), while onboard hardware is purchased for every registered vehicle whether driving or parked, so a low-utilization fleet amortizes a small shared GPU pool across a large registered base.

\begin{figure}[width=0.99\textwidth, pos = t]
\centering
\includegraphics[width=0.8\linewidth]{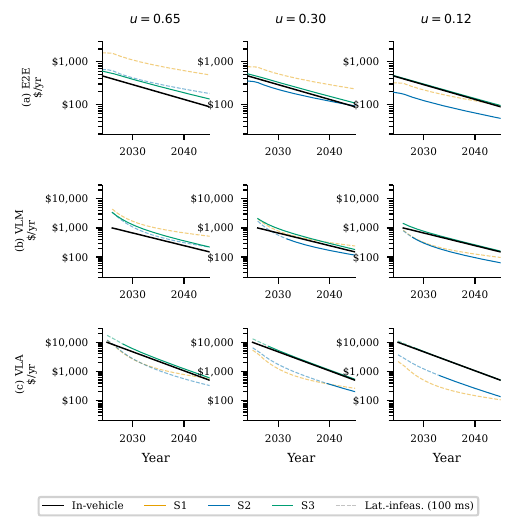}
\caption{Regime 3 (cost-bound), reactive budget. NYC per-vehicle hybrid cost versus the in-vehicle baseline (black) over time at 5G-Advanced. Rows: model classes. Columns: utilization. Solid curves are latency-feasible under the 100~ms budget; dashed segments are latency-infeasible. VLA cloud (S2) falls below the \$8{,}500 baseline once it clears the compute regime.}
\label{fig:nyc_cost_crossover}
\end{figure}

Figure~\ref{fig:nyc_cost_crossover} tracks per-vehicle hybrid cost (cloud plus residual onboard) against the in-vehicle baseline over time, under the reactive budget. The latency gate is visible directly: VLA branches are dashed (infeasible) in the near term and become solid only once the compute regime clears, after which feature-level VLA cloud falls below the \$8{,}500 onboard baseline. The sharing effect is ordered by model class. It is weak for E2E, whose \$400 onboard baseline is already cheap; intermediate for VLM; and strongest for VLA, whose costly, intermittently used onboard hardware is exactly what shared inference displaces. Within these cells---10\% penetration at moderate-to-low utilization---S2 is the cheapest cloud strategy at every crossover, because S1's network cost and S3's residual onboard hardware both erode the saving. Across the full penetration--utilization grid the pattern is strong but not universal: S1 can win at very low penetration, where active demand is too small for its network cost to matter, and S3 wins the cells where S2 is not yet latency-admissible.

\begin{figure}[width=\textwidth, pos = t]
\centering
\includegraphics[width=\linewidth]{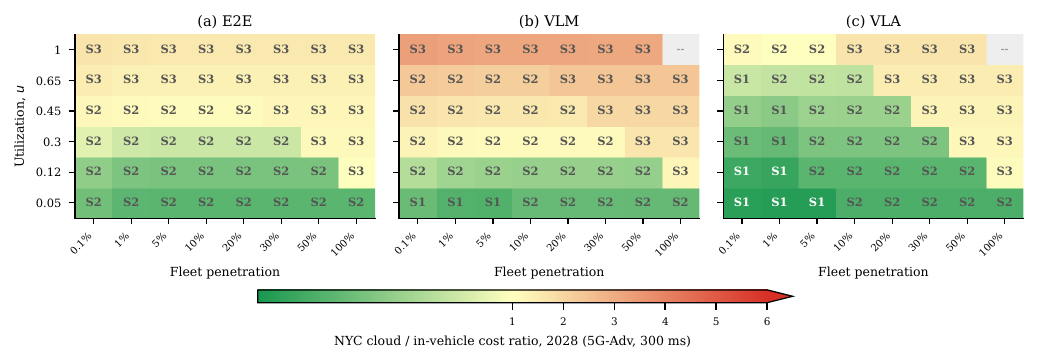}
\caption{Regime 3 (cost-bound), deliberative tier in 2028, where VLA is latency-admissible. Cheapest feasible NYC cloud cost divided by the in-vehicle baseline; cell text gives the cheapest feasible strategy and gray cells are infeasible. Green marks cloud-cheaper cells. For VLA the cost-attractive region is large and dominated by S2.}
\label{fig:nyc_cost_ratio}
\end{figure}

Figure~\ref{fig:nyc_cost_ratio} isolates the economics from the latency gate by fixing the deliberative tier in 2028, where VLA is admissible. The choice of 5G-Advanced and 2028 is deliberate rather than incidental. 5G-Advanced is the \emph{threshold} generation for VLA feature-level offloading (Section~\ref{sec:cs_comm}): 5G fails the dense-corridor uplink outright, whereas 6G supplies so much headroom that raw-sensor offloading (S1)---which keeps the least residual hardware onboard---becomes feasible and trivially cheapest, masking the very strategy trade-off this figure is meant to expose. Only at 5G-Advanced does the uplink still gate S1 in dense cells, so the cheapest \emph{feasible} strategy is genuinely contested and, as the panels show, won by S2. The year 2028 is chosen as a near-term and deliberately conservative snapshot: it is roughly when VLA both reaches market and clears the deliberative budget, and because cloud GPU prices only decline afterward it \emph{understates} rather than flatters cloud attractiveness. The deliberative tier is shown because, under the reactive budget, 5G-Advanced admits no VLA-S2 cell at any year in this grid, so it is the only tier in which 5G-Advanced VLA economics can be evaluated at all. The VLA panel is broadly cost-attractive (green) at low-to-moderate utilization, and the cheapest feasible strategy in that region is overwhelmingly S2. The contrast with E2E is the point: the E2E panel is mostly cost-neutral or unattractive because the onboard pipeline is already inexpensive, whereas the VLA panel rewards offloading because the avoided hardware is expensive. Reading the strategy label and color together shows that for VLA, S2 is not merely the branch that passes communication---it is the branch that preserves enough hardware saving to beat the baseline. Under the reactive budget the same economics hold where S2 is latency-admissible, but the tight budget makes VLA-S2 infeasible across much of the grid, so the cheapest \emph{feasible} strategy there shifts to S3, which keeps more compute onboard and rarely beats the baseline. The clean S2-dominated crossover is thus a property of the deliberative tier and the low-penetration reactive slice. One caveat applies to the deliberative-tier economics: the cost ratios charge only each strategy's split residual and exclude the onboard reactive-fallback hardware a deliberative branch must carry (Section~\ref{sec:discussion}), so the VLA cost-attractive region shown is an optimistic bound.

\textbf{Once feasibility is met, utilization and GPU economics---not communication generation---determine whether cloud offloading is cheaper than full onboard deployment, and feature-level offloading (S2) is where the VLA cost advantage concentrates.}

\section{Discussion}
\label{sec:discussion}

\subsection{Synthesis}

The results challenge two simplified narratives about cloud-based autonomous driving.
The 6G literature often frames next-generation communication as the main enabler, while the AV industry often treats full in-vehicle compute as the only deployable architecture.
Both views miss that three different constraints bind in sequence, and which one binds depends on the model class, the GPU year, and the latency budget.

First, in dense cells the uplink binds: S1 is eliminated early, and S2 becomes practical only when the access network can carry feature-level traffic under realistic loading.
Second, under the reactive 100~ms budget the cloud inference floor binds for VLA: autoregressive decode is memory-bandwidth-bound, so near-term VLA is latency-infeasible regardless of bandwidth or GPU cost, and becomes admissible only as memory bandwidth grows (sooner on 6G, later on 5G-Advanced).
The 300~ms deliberative tier relaxes this, but presumes an onboard reactive controller, so it does not eliminate onboard hardware---it reclassifies the cloud as a non-safety-critical planning aid.
Third, once a branch is admissible, utilization binds the economics: because cloud GPUs are provisioned for concurrent demand while onboard hardware is paid for per vehicle, low-utilization fleets gain the most, and the gain is largest for VLA, whose peak hardware is expensive and usually idle.

Feature-level offloading (S2) is the common solution across all three regimes because it balances the failure modes at the ends of the spectrum.
S1 minimizes residual in-vehicle computation but uses too much uplink capacity in dense cells.
S3 minimizes uplink traffic but keeps enough hardware in the vehicle to reduce the economic value of offloading.
S2 partitions the pipeline at a natural boundary: the vision backbone stays local, while compute-heavy prediction, planning, and language components move to shared infrastructure.
S2 is therefore not a universal optimum, but it is the most robust middle point---and where it is latency-admissible, it is also where the VLA cost crossover concentrates.

\subsection{Policy Implications}

These findings translate into policy choices because cloud driving depends on communication deployment, shared edge infrastructure, safety certification, and utilization-aware capacity planning rather than on vehicle hardware alone.

\textbf{Scenario-specific communication migration.} Urban planners should connect communication upgrades to expected AV penetration and utilization rather than to a generic 6G timeline. In dense city corridors, the NYC sweep suggests that plain 5G has too little headroom for S1 and can also fail moderate S2 deployment as utilization increases. Here, headroom is the spare capacity left after current demand is served. 5G-Advanced should therefore be treated as the minimum migration target for early feature-level offloading, while 6G becomes important once penetration and utilization move beyond moderate adoption.

\textbf{Uplink-heavy spectrum allocation for AV corridors.} Current cellular networks are designed mainly for downlink-heavy consumer traffic such as video streaming and web browsing. Cloud-based AV traffic reverses that pattern: most bandwidth demand is uplink, and the downlink carries only compact trajectory commands. Without uplink-aware spectrum allocation, scheduler tuning, and interference management in AV-dense corridors, practical deployments can bottleneck before reaching the theoretical cell capacity. This is another reason S2 should be the planning reference: it keeps uplink demand low enough to be plausible while still revealing the communication investments dense corridors require.

\textbf{Public-private co-investment in shared edge infrastructure.} The facility location optimization indicates that placing GPU clusters at existing cell towers captures much of the edge-computing cost benefit while avoiding redundant infrastructure build-out. A single AV operator may not be able to justify a full metropolitan edge build-out, but a shared infrastructure model, analogous to tower sharing among mobile carriers, can amortize facility costs across operators. Outside dense urban cores, the shared-infrastructure priority shifts from communication generation alone to GPU economics. Once feasibility is satisfied, the public-interest lever is not only more spectrum but better sharing of expensive cloud hardware.

\textbf{Safety certification for cloud-dependent architectures.} Current NHTSA guidelines~\citep{nhtsa2017ads, usdot2020av40} implicitly assume self-contained vehicles. Cloud-dependent architectures add failure modes that those frameworks do not yet cover, including network outage, edge-site failure, and latency spikes. We recommend certification requirements for mandatory in-vehicle fallback capability, at minimum S3-level compressed offloading or full in-vehicle E2E inference, so loss of cloud connectivity degrades performance gracefully rather than catastrophically. This requirement is sharper under the two-budget view: any branch that relies on the 300~ms deliberative tier is, by construction, conditioned on an onboard reactive controller that closes the 100~ms loop locally, so deliberative cloud planning can supplement but never replace in-vehicle reactive autonomy. The offloading spectrum can serve as a certification ladder: S2 may be the economic mainline, but S3 or full local fallback should remain available as the safety backstop.

\textbf{Utilization-aware infrastructure sizing.} Edge capacity should be sized for peak \textit{concurrent} demand rather than for total registered fleet size. At $u=0.12$, only 12\% of the fleet is active at the same time; sizing for $N$ instead of $N \cdot u$ would over-provision GPU capacity by several factors, increasing cost and weakening the case for cloud offloading. This utilization effect is why the cloud advantage appears first in low- and moderate-utilization branches. After communication feasibility is met, utilization is the main planning variable because it determines whether shared cloud GPUs replace idle onboard hardware or merely duplicate it.

\subsection{Limitations}

\textbf{Model architecture evolution.} Techniques such as speculative decoding~\citep{leviathan2023speculative}, mixture-of-experts~\citep{fedus2022switch}, and model distillation~\citep{hinton2015distilling} could change the compute required for each inference. Speculative decoding tries to generate tokens faster by proposing and verifying them, while model distillation trains a smaller model to reproduce behavior from a larger one. The framework can absorb new architectures by updating TFLOPS and decode-token inputs, but the specific crossover years depend on current model classes. A large reduction in decoder-side cost could make VLM offloading more attractive or move the best split point away from today's S2 assumption.

\textbf{6G parameters.} The 6G values use ITU IMT-2030 targets~\citep{itu2023imt2030}, which are aspirational rather than guaranteed deployment values. If realized capacity is lower, the urban feasibility boundary would move, and some high-penetration S2 cells would need denser infrastructure than our sweep indicates. The qualitative result is robust to moderate parameter changes: communication generation binds at urban density, but it becomes less decisive after branches pass feasibility. The precise migration point from 5G-Advanced to 6G remains uncertain.

\textbf{Static utilization.} We treat utilization $u$ as constant and do not model spatiotemporal demand variation. Spatiotemporal variation refers to demand changing across both location and time, for example a downtown corridor during rush hour versus a suburban road at night. In practice, peak-demand cells may see 3--5$\times$ the average loading, so the analysis may understate hotspot difficulty and overstate difficulty in quieter areas. This matters most for the urban results: local hotspot congestion could require corridor-specific upgrades even when citywide average penetration still appears feasible.

\textbf{Energy costs.} GPU energy is excluded from the TCO. At current U.S. rates, energy would add about 5--12\% to cloud costs, shifting crossovers 1--2 years later for marginal cases without changing the qualitative results. Marginal cases are branches close to cost parity between cloud and onboard deployment. Energy therefore matters most where the cloud case is already weak, making it more likely to narrow the economically feasible region than to change the preferred strategy.

\textbf{Roofline model.} The roofline model assumes a clean separation between compute-bound and bandwidth-bound inference, calibrated against one hardware/software stack: Blackwell with TensorRT-LLM~\citep{nvidia2026alpamayo}. It also treats VLA decoding as fully autoregressive, which is a conservative upper bound on the compute-bound regime: diffusion- or flow-based action decoders generate trajectory tokens in parallel rather than token-by-token, cutting decode latency substantially (to the tens of milliseconds) and moving VLA's reactive-budget admissibility several years earlier. Likewise, other serving frameworks, batching policies, speculative decoding, or FP8 instead of our FP16 baseline could shift the phase split and reduce VLA cloud cost and latency. Future serving and decoder improvements may therefore strengthen the cloud case and soften the compute-bound regime, even if they do not change the communication ordering among S1, S2, and S3.

\textbf{GPU evolution.} The decelerating growth model is calibrated to four data points from A100 to B300. A slowdown in GPU gains would delay VLM crossover and reduce the VLA cost advantage; conversely, accelerator architecture improvements could keep growth above the 10\%/yr long-run floor. The exact crossover year is therefore uncertain. The broader interpretation is more stable: once communication is feasible, utilization and cloud accelerator productivity remain the dominant economic levers.

\section{Conclusion}
\label{sec:conclusion}

\textbf{Can the cloud drive? Conditionally, and in a specific order.}
Cloud offloading is not a complete replacement for onboard compute, but it can be markedly more cost-effective once the network and latency conditions are met. Its advantage is hardware utilization: cloud GPUs are allocated only to active vehicles, whereas onboard hardware is purchased for every car whether driving or parked.

A single New York City case study, read under a reactive 100~ms and a deliberative 300~ms budget, shows that three constraints bind in sequence rather than one binding everywhere.
Communication binds first in dense cells: S1 is too uplink-heavy, many cells require 6G, and S2 fits within the 5G-Advanced/6G envelope.
Compute binds next under the reactive budget: VLA inference is memory-bandwidth-bound at the decode stage, so near-term VLA is latency-infeasible regardless of bandwidth and becomes admissible only as memory bandwidth grows---its floor clears the reactive budget around 2027, after which 6G admits feature-level VLA by $\sim$2028 while 5G-Advanced admits it only at light cell loading and not at the dense reference corridor within the horizon---while the deliberative tier admits it from 2026 but only behind an onboard reactive fallback.
Cost binds last: once admissible, utilization and GPU productivity decide the comparison, and the advantage is largest for VLA, whose expensive, often-idle onboard hardware is what shared inference displaces.

Two takeaways follow. Architecturally, feature-level offloading (S2) keeps perception near the sensors and offloads the cloud-worthy prediction and planning stages; it is the most robust middle point and where the VLA cost crossover concentrates. Methodologically, latency decides which model class is admissible in which year, and cost decides whether it is economical---so the answer to ``can the cloud drive?'' is not binary but a function of model class, GPU year, and the latency budget the deployment must meet.

\printcredits




\section*{Declaration of generative AI and AI-assisted technologies in the manuscript preparation process}
During the preparation of this work the authors used a generative AI coding assistant to support code development, figure editing, and language editing. After using this tool, the authors reviewed and edited the content as needed and take full responsibility for the content of the publication.

\bibliographystyle{elsarticle-num}
\bibliography{references}

\end{document}